\documentclass[aps,pra,twocolumn,letterpaper,showpacs]{revtex4-1}

\usepackage{amsmath}
\usepackage{amssymb}
\usepackage{amsfonts}
\usepackage{txfonts}
\usepackage{bbold}
\usepackage{color}
\usepackage{bm}
\usepackage{graphicx}
\usepackage{ulem}
\usepackage{hyperref}
\newcommand {\eref}[1]{(\ref{#1})}

\newcommand{\x}{{\mathbf x}}

\newcommand{\bk}{{\mathbf k}}

\newcommand{\brr}{{\mathbf r}}

\newcommand{\ecut}{\epsilon_{\rm{cut}}}

\newcommand{\EQ}[1]{\begin{eqnarray}#1\end{eqnarray}}

\newcommand{\vcore}{b}
\newcommand{\vslope}{\Lambda}

\newcommand{\pxa}{\mathbf{r},t }
\newcommand{\pxb}{ \mathbf{r} }

\newcommand{\py}{\chi}   
\newcommand{\pxc}{{\sigma}}
\newcommand{\pz}{k\xi}

\begin{document}
\normalem

\title{Energy spectra of vortex distributions in two-dimensional quantum turbulence}
\author{Ashton S. Bradley} 
\address{Jack Dodd Center for Quantum Technology, Department of Physics, University of Otago, Dunedin 9016, New Zealand.}
\author{Brian P. Anderson}
\address{College of Optical Sciences, University of Arizona, Tucson, Arizona 85721, USA.}

\begin{abstract}

We theoretically explore key concepts of two-dimensional turbulence in a homogeneous  compressible superfluid described by a dissipative two-dimensional Gross-Pitaeveskii equation. Such a fluid supports quantized vortices that have a size characterized by the healing length $\xi$. We show that for the divergence-free portion of the superfluid velocity field, the kinetic energy spectrum over wavenumber $k$ may be decomposed into an ultraviolet regime ($k\gg \xi^{-1}$) having a universal $k^{-3}$ scaling arising from the vortex core structure, and an infrared regime ($k\ll\xi^{-1}$) with a spectrum that arises purely from the configuration of the vortices. The Novikov power-law distribution of intervortex distances with exponent $-1/3$ for vortices of the same sign of circulation leads to an infrared kinetic energy spectrum with a Kolmogorov $k^{-5/3}$ power law, consistent with the existence of an inertial range.  The presence of these $k^{-3}$ and $k^{-5/3}$ power laws, together with
the constraint of continuity at the smallest configurational scale $k\approx\xi^{-1}$, allows us to derive a new analytical expression for the Kolmogorov constant that we test against a numerical simulation of a forced homogeneous compressible two-dimensional superfluid. The numerical simulation corroborates our analysis of the spectral features of the kinetic energy distribution, once we introduce the concept of a {\em clustered fraction} consisting of the fraction of vortices that have the same sign of circulation as their nearest neighboring vortices. Our analysis presents a new approach to understanding two-dimensional quantum turbulence and interpreting similarities and differences with classical two-dimensional turbulence, and suggests new methods to characterize vortex turbulence in two-dimensional quantum fluids via vortex position and circulation measurements.
\end{abstract}
\maketitle
\section{Introduction}
Turbulence in three-dimensional (3D) classical fluids is associated with a cascade of energy from large length scales defined by the details of an energy-forcing mechanism, to small length scales where viscous damping removes kinetic energy from the fluid.  This range of length scales, and the range of associated wavenumbers $k$, define the \emph{inertial range} of energy flux~\cite{Les2008.Turbulence}.  As shown by Kolmogorov in 1941~\cite{Kolmogorov1941}, the energy cascade corresponds to a kinetic energy spectrum that is proportional to $k^{-5/3}$ in the inertial range.   Turbulence in a 3D fluid is also often associated with the decay of large patches of vorticity into ever smaller regions of vorticity; this 
\emph{Richardson cascade} of vorticity provides an important visual picture of the fluid dynamics involved in 3D turbulence~\cite{Sreenivasan1999a}.

Remarkably, two-dimensional (2D) incompressible classical fluids exhibit very different turbulent flow characteristics due to the existence of an additional inviscid invariant: in the absence of forcing and dissipation, the 
total {\em enstrophy} \cite{whatisenstrophy} of a 2D fluid is conserved in addition to the fluid's kinetic energy~\cite{Kra1967.PF10.1417,Lei1968.PF11.671,Bat1969.PF12.II233,Kra1980.RPP5.547,Boffetta12a}.  The fluid dynamics during forced 2D turbulence are highly distinctive when compared with 3D flows:  rather than decaying into smaller patches, vorticity aggregates into larger coherent rotating structures~\cite{Kellay2002a} (see \cite{Chen06a} for a more detailed picture in terms of turbulent stress imposed on small-scale vortices).  Accompanying these 2D fluid dynamics is an \emph{inverse} energy cascade in which energy moves from a small forcing scale to progressively larger length scales, defining an inertial range for energy transport with energy flux in a direction opposite that of 3D turbulence.  Eventually, energy is transported into flows characterized by length scales that are on the order the system size~\cite{Mon1974.PF6.1139}, for which dissipation may occur.  Additionally, there is an enstrophy cascade, in which enstrophy is transported from the forcing scale to progressively smaller scales~\cite{Rutgers1998a}.  Thus in 2D turbulence, the kinetic energy distribution contains at least these two distinctly different spectral regimes.

{\em Quantum} turbulence~\cite{Vinen07a} involves chaotic flow in a superfluid~\cite{Bar2001.book.Superfluid,Berloff02a,Tsubota09a,Kozik09a,Nowak2011a} and is often associated with a random vortex tangle in 3D~\cite{Bar2001.book.Superfluid}. In general, the quantization of circulation strongly constrains the velocity fields allowed in quantum turbulence, which must be irrotational everywhere within the fluid, yet inertial ranges with $k^{-5/3}$ spectral dependence are still found in 3D quantum turbulence~\cite{Maurer1998a}. In an {\em incompressible} superfluid (such as HeII), the vortex core diameter can be neglected for all practical purposes, inspiring the study of point-vortex models of superfluid dynamics. Such a model was used by Onsager to first predict the aggregation of vortices within inviscid 2D fluids, and was the context for his prediction of the quantization of vortex circulation in a superfluid~\cite{Ons1949.NC6s2.279}. Despite the historical importance of this approach in stimulating advances in 2D classical turbulence \cite{Eyi2006.RMP78.87}, characteristics of 2DQT remain little known, due in part to the difficulty of achieving the necessary 2D confinement for incompressible superfluids. The increasing relevance of 3D turbulence concepts to dilute-gas Bose-Einstein condensate (BEC) experiments~\cite{Henn09a,Henn2010a,Seman2011a} and recent theoretical work on 2DQT~\cite{Parker05a,Nazarenko07a,Horng09a,Numasato10a,Numasato10b,Nowak2011a,Schole2012a} have highlighted the need for a treatment of turbulence in 2D superfluid systems that incorporates the concept of compressibility from the outset.  Motivated by recent experimental demonstrations of the confinement needed to study 2DQT in a dilute-gas BECs \cite{Neely10a, Neely2012a} our aim in the present paper is to present a new approach to solving some of the open problems of 2DQT in the context of such a system. 

In a BEC, the vortex core size is non-negligible, and stems from the healing length $\xi$, a scale of fundamental importance in BEC dynamics that is typically about two orders of magnitude smaller than the system size~\cite{Dalfovo1999}. Compressibility also allows for a rich array of physical phenomena in these superfluids; in particular, a vortex dipole \cite{Neely10a} (comprised of two vortices of opposite sign of circulation) can recombine, releasing vortex energy as a burst of acoustic waves.  The opposite process of vortex dipole generation from sound may also efficiently occur. 

Recent theoretical studies of decaying quantum turbulence in 2D BECs have shown that when the vortex dipole annihilation process is dominant it sets up a direct cascade of energy over the scales associated with the dipole decay, suggesting that this annihilation mechanism could prohibit an inverse energy cascade from occurring in a compressible superfluid~\cite{Numasato10a,Numasato10b}. Moreover,  enstrophy in a quantum fluid is associated with the number of vortex cores; if vortices annihilate, then enstrophy may not be  conserved, bringing into question the existence of energy and enstrophy cascades in 2DQT, and the universal nature of 2D turbulence and its correspondence with 2DQT.   

The general characteristics of 2D quantum turbulence in compressible quantum fluids, including the capacity for these systems to show an inverse energy cascade, enstrophy conservation, and vortex aggregation have thus remained largely unknown. However, a recent study of the formation of vortex dipoles during the breakdown of superfluid flow around an obstacle in a highly oblate BEC experimentally and numerically observed aggregation of like-sign vortices into larger-scale coherent structures~\cite{Neely10a}, and found time scales over which vortex number and hence enstrophy may remain constant. The vortex {\em clustering} effect inhibits the dipole-decay mechanism by keeping vortices distant from antivortices (vortices of opposite circulation), and suggests that an inverse cascade might be observed under the right conditions of forcing.  System dynamics consistent with the existence of an inverse energy cascade were indeed found in a recent study of forced 2DQT in a BEC \cite{Neely2012a}.  

In this article we address 2D quantum turbulence in a compressible quantum fluid from an analytical perspective. We determine the kinetic energy spectra of vortex distributions in a homogeneous compressible superfluid in a quasi-exact manner via an analytical treatment of the physics of the vortex core. We are thus able to study the properties of vortex configurations and their resulting spectra in BEC. We develop a technique to sample spatially localized vortex distributions with power-law behavior over a well defined scale range. We are thus able to identify the conditions for an inertial range in fully polarized and neutral systems. A polarized cluster is sampled using a specific exponent for the vortex locations relative to the cluster center, which is size and scale dependent. The specific radial exponent is shown to determine the velocity distribution in the classical limit and we thus identify an expanding inertial range with a steepening velocity distribution. 
\par
Making use of the universality of the spectral region generated by the vortex core, we identify an analytical form of the Kolmogorov constant that we test against dynamical simulations of the damped GPE. The derivation of the Kolmogorov constant occurs for a highly idealized vortex distribution. Thus the complex flows generated by real forcing require that we introduce a new parameter called the {\em clustered fraction}, and modify our ansatz to account for imperfect clustering, based on the universality of the Kolmogorov constant. The modified ansatz agrees well with the numerical simulations of grid turbulence, supporting our analytical identification of the Kolmogorov constant.

\section{Background}
The starting point for much of BEC theory is the Gross-Pitaevskii equation (GPE), which provides a capable description of trapped Bose-Einstein condensates at zero temperature \cite{Dalfovo1999}. Our model, outlined below, consists of a damped GPE (dGPE) description of a finite-temperature BEC which can be derived from the stochastic GPE theory~\cite{SGPEII,Bradley08a}. In this section we develop a link between the dGPE and the classical Navier-Stokes equation, identifying a {\em quantum viscosity} arising from the damping. The corresponding Reynolds number is defined in direct analogy with classical fluids. We then state some key properties of a single quantum vortex, and define the decomposition of kinetic energy into its compressible and incompressible components.
\subsection{Damped Gross-Pitaevskii theory}
The damped Gross-Pitaevskii equation of motion for the quantum fluid wave function $\psi(\pxa)$ has been obtained phenomenologically~\cite{Choi1998}, within ZNG theory~\cite{Tsubota2002}, and via a microscopic reservoir theory~\cite{Penckwitt2002,SGPEI,SGPEII}, and we will consider it within the context of the latter framework, for which the full equation of motion is the Stochastic Projected Gross-Pitaevskii equation (SPGPE). The SPGPE is derived by treating all atoms with energies above an appropriately chosen energy cutoff $\ecut$ as thermalized, leading to a grand-canonical description of the atoms below $\ecut$.  A dimensionless temperature-dependent rate $\gamma$ describes Bose-enhanced collisions between thermal reservoir atoms and atoms in the BEC.  Neglecting the noise, we obtain the equation of motion for the condensate wave function (in the frame rotating with the chemical potential $\mu$)
\begin{equation}\label{eom}
i\hbar\frac{\partial \psi(\pxa)}{\partial t}=(i\gamma-1)(\mu-{\cal L})\psi(\pxa).
\end{equation} 
For atoms of mass $m$ in an external potential $V(\pxa)$, the operator ${\cal L}$ gives the GPE evolution:
\begin{equation}
{\cal L}\psi(\pxa)\equiv \left(-\frac{\hbar^2}{2m}\nabla^2+V(\pxa)+g|\psi(\pxa)|^2\right)\psi(\pxa),
\end{equation}
where the interaction parameter is $g=4\pi\hbar^2 a/m$, for $s$-wave scattering length $a$. This equation of motion has been used extensively in previous studies of vortex dynamics~\cite{Tsubota2002,Penckwitt2002,Madarassy08a,Rooney10a} and provides a capable description of dynamical BEC phenomena. In general the damping parameter is small ($\gamma\ll 1$), and it is typically much smaller than any other rates characterizing the  evolution.   

Defining the Gross-Pitaevskii Hamiltonian 
\begin{equation}\label{hc}
H_{\rm C}=\mint{^3\pxb}\,\left\{\frac{\hbar^2}{2m}|\nabla\psi(\pxa)|^2+V(\pxa)|\psi(\pxa)|^2+\frac{g}{2}|\psi(\pxa)|^4\right\},
\end{equation}
and condensate atom number
\begin{equation}
N_{\rm C}=\mint{^3\pxb}|\psi(\pxa)|^2,
\end{equation}
the equation of motion \eref{eom} evolves the grand-canonical Hamiltonian $K_{\rm C}=H_{\rm C}-\mu N_{\rm C}$ according to
\EQ{
\frac{dK_{\rm C}}{dt}=-\frac{2\gamma}{\hbar}\mint{^3\pxb}|(\mu-{\cal L})\psi(\pxa)|^2.
}
The stationary solution minimizing $K_{\rm C}$ is the ground state satisfying $\mu\psi_0(\mathbf{r})\equiv{\cal L}\psi_0(\mathbf{r})$. This is a consequence of the nonlinear form of the damping in \eref{eom}. The damping term arises from collisions between high energy atoms that lead to a Bose-enhanced growth of the matter wave field, with instantaneous energy determined by ${\cal L}$. The equation of motion thus describes a system coupled to a thermal reservoir in the chosen frame of reference.

The SPGPE provides a rigorous framework for the dGPE derivation, originating from a microscopic treatment of the reservoir interaction. In particular, $\gamma$ can be calculated explicitly~\cite{Bradley08a} for a system with well-defined reservoir parameters $\mu, T,$ and $\ecut$, i.e. a system close to thermal equilibrium. In essence it is computed via a reduced Boltzmann collision integral that accounts for all irreversible $s$-wave interactions that can change the condensate population by interacting with the thermal cloud. If the thermal cloud is 3D (i.e.  $\beta^{-1} \equiv k_BT$ is greater than the potential well mode spacing in each spatial dimension) the damping takes the explicit form
\EQ{\label{damping}
\gamma=\gamma_0\sum_{j=1}^\infty \frac{e^{\beta\mu(j+1)}}{e^{2\beta j\epsilon_{\rm cut} }}\Phi\left[\frac{e^{\beta\mu}}{e^{\beta j \epsilon_{\rm cut} }},1,j\right]^2,
}
where $\Phi[z,s,\alpha]$ is the Lerch transcendent, and
\EQ{\label{gam0}
\gamma_0=8a^2/\lambda_{dB}^2,
}
with $\lambda_{dB}\equiv\sqrt{2\pi\hbar^2/m k_BT}$ the thermal deBroglie wavelength. 
The dimensionless rate $\gamma_0$ provides a useful estimate of the full damping strength when the cutoff $\ecut$ is unknown. Equation \eref{damping} is independent of position, and valid over the region $V(\pxa)\leq 2\epsilon_{\rm cut}/3$, provided the potential can be treated semi-classically~\cite{Bradley08a}. The summation gives Bose-enhancement corrections due to the Bose-Einstein distributed reservoir atoms, and is typically of order 1-20 in SPGPE simulations with a consistently determined energy cutoff~\cite{Blakie08a}. Typicaly $\gamma \sim 5\times 10^{-4}$ in $^{87}$Rb experiments~\cite{Rooney10a,Rooney11a}. 

\subsection{Heuristic derivation of a quantum Reynolds number}
In this section we consider the role of dissipation within the dGPE description, and show how to recover the celebrated Navier-Stokes equation (NSE). In doing so we find an explicit expression for the viscosity which has a microscopic {\em quantum} origin, stemming from s-wave scattering of incoherent reservoir particles with a coherent superfluid. 
While not offering a practical reformulation (the GPE and its generalizations are capable numerical workhorses), this indicates a connection between the dGPE and the NSE in the hydrodynamic regime, allowing the identification of a parameter analogous to the kinematic viscosity  of classical fluids. 

The fluid dynamics interpretation of the Gross-Pitaevskii equation is based on the Madelung transformation, which we now apply to the damped GPE \eref{eom}, writing $\psi(\pxb,t)=\sqrt{\rho(\pxb,t)}\exp{[i\Theta(\pxb,t)]}$, where $\rho(\pxb,t)$ is the number density of the superfluid (number of atoms per unit volume), and $\Theta(\pxb,t)$ is the macroscopic phase of the quantum fluid. The velocity is then given by $\mathbf{v(\pxb},t)=\hbar\nabla\Theta(\pxb,t)/m$. The resulting equations of motion (with implicit $t$ and $\textbf{r}$ dependence)  for density and velocity are then given by
\begin{eqnarray}\label{cont}
\frac{\partial \rho}{\partial t}+\nabla\cdot(\rho \mathbf{v})&=&\frac{2\rho \gamma}{\hbar} (\mu-U_{\rm eff}),\\\label{Euler}
m\frac{\partial \mathbf{v}}{\partial t}&=&-\nabla\left(U_{\rm eff}-\frac{\hbar \gamma}{2\rho}\nabla\cdot(\rho\mathbf{v})\right),
\end{eqnarray}
where an effective potential $U_{\rm eff}$ is defined as 
\begin{equation}\label{Ueff}
U_{\rm eff}(\pxb,t)=\frac{m \mathbf{v}^2}{2}+V+g\rho-\frac{\hbar^2}{2m}\frac{\nabla^2\sqrt{\rho}}{\sqrt{\rho}}.
\end{equation}
The last term is called the {\em quantum pressure}, which is very small except where $\rho$ changes sharply, such as near vortex cores. By neglecting this term in the absence of dissipation we are considering the so-called {\em hydrodynamic regime}.

We now consider the $\gamma$ term in \eref{Euler} 
\begin{equation}\label{Gamterm}
\frac{\hbar \gamma}{2}\nabla \left(\frac{1}{\rho}\nabla\cdot(\rho\mathbf{v})\right)=\frac{\hbar \gamma}{2}\left(\nabla(\nabla\cdot\mathbf{v})+\nabla\frac{\mathbf{v}\cdot\nabla\rho}{\rho}\right).
\end{equation}
Note that $\mathbf{v}\cdot\nabla\rho\equiv 0$ for an isolated quantum vortex. In the absence of acoustic energy this will also be a good approximation for a system of vortices provided their cores are well separated, since the density gradient of each vortex is localized to a region where the velocity is dominated by the single-vortex velocity field. It should thus be a reasonable approximation to neglect the second term in \eref{Gamterm}. In a superfluid the curl term in the expansion $\nabla(\nabla\cdot\mathbf{v})=\nabla\times(\nabla\times\mathbf{v})+\nabla^2\mathbf{v}$ may also be consistently neglected away from vortex cores; similarly we neglect the curl term in $\nabla(\mathbf{v}\cdot\mathbf{v})=2\,(\mathbf{v}\cdot\nabla)\mathbf{v}+2\,\mathbf{v}\times(\nabla\times\mathbf{v})$ when taking the gradient of \eref{Ueff}. We then find that \eref{Euler} reduces to a quantum Navier-Stokes equation for the velocity field: 
\begin{equation}\label{qnsdef}
\frac{\partial \mathbf{v}}{\partial t}+(\mathbf{v}\cdot\nabla)\mathbf{v}=-\frac{1}{m}\nabla\left(V+g\rho\right)+\nu_q\nabla^2\mathbf{v}
\end{equation}
where the {\em kinematic quantum viscosity} is 
\begin{equation}
\nu_q\equiv\frac{\hbar\gamma}{2m},
\end{equation}
in analogy with classical fluids.
In this regime, \eref{qnsdef} is coupled to the continuity equation
\EQ{\label{cont}
\frac{\partial \rho}{\partial t}+\nabla\cdot(\rho \mathbf{v})&=&\frac{2\rho \gamma}{\hbar} (\mu-U_H)
}
with hydrodynamic potential
\begin{equation}\label{UH}
U_H(\pxb)\equiv\frac{m \mathbf{v}^2}{2}+V+g\rho.
\end{equation}
The source term in \eref{cont} drives the system towards particle-number equilibrium with the reservoir. In the Thomas-Fermi regime $\mu-U_H(\pxb)\approx 0$, restoring approximate particle-number conservation.

Making use of \eref{gam0}, we can give an order-of-magnitude estimate for the viscosity
\EQ{\label{nuqest}
\nu_q^0\equiv\frac{\hbar\gamma_0}{2m}=\frac{2a^2 k_BT}{\pi\hbar}.
}
We can also estimate a quantum Reynolds number as
\EQ{\label{Ren}
Re_q^0\equiv\frac{UL}{\nu_q^0}=\frac{\pi\hbar}{2a^2 }\frac{UL}{k_BT}
}
for BEC flow with characteristic speed $U$ and length scale $L$. We can write the quantum Reynolds number as
\EQ{\label{Re0}
Re_q^0=\frac{\lambda_{dB}^2}{a^2 }\frac{mUL}{4\hbar}.
}
We note that temperature only enters the expression through the deBroglie wavelength of the matter wave field, in the ratio $\lambda_{dB}/a $, which is typically very large for a BEC. Note that strong scattering corresponds to strong damping, and hence low Reynolds number. A large deBroglie wavelength corresponds to a relatively cold system, which is hence expected to be weakly damped and have a high Reynolds number. 

We thus have a dimensionless ratio in the form
\EQ{\label{Redimless}
Re_q^0\sim\frac{[\mbox{\small deBroglie wavelength}]^2}{[\mbox{\small scattering length}]^2}\cdot\frac{[\mbox{\small flow momentum}]}{[\mbox{\small quantum momentum}]}\;\;\;
}
where $\hbar/L$ is interpreted as the quantum momentum associated with the transverse length scale of the flow.

A concrete example is provided by a recent experimental study of 2DQT generated by stirring a highly oblate, toroidally confined BEC~\cite{Neely2012a}. The initial system consists of $\sim 2.6\times 10^6$ atoms of $^{87}$Rb at a temperature of $\sim 100{\rm nK}$. Using these numbers in our analysis the dimensionless damping parameter $\gamma_0\sim 6\times 10^{-4}$ gives a kinematic quantum viscosity $\nu_q^0\sim6\times 10^{-2}\mu{\rm m}^2\,{\rm s}^{-1}$. The trapping potential confines the flow to an annular channel of width $L\sim 30\,\mu{\rm m}$. The nominal flow speed can be estimated from numerical simulations of the dGPE~\cite{Neely2012a}, giving a value $U\sim 5\, \mu{\rm m\, s}^{-1}$ as the peak value occurring in the bulk flow during the stirring sequence. These values give an estimate $Re_q^0\sim 600$.  We can alternatively estimate a Reynolds number at the scale of the forcing in the experiment, which is of order of the size of the vortex dipoles nucleated, $d\sim 10\,\xi$, with healing length $\xi\sim 0.5\,\mu{\rm m}$.  Such dipoles have a characteristic speed ${\rm v}_d\sim 146\,\mu{\rm ms}^{-1}$, for which we estimate the Reynolds number of the forcing scale as $\sim {\rm v}_dd/\nu_q^0=6.2\times 10^3$. These large values suggest that turbulent flow in a finite-temperature BEC may exist across a wide range of length scales if the analogy is made with the classical Reynolds numbers that correspond to turbulent flow~\cite{Les2008.Turbulence}. This interpretation is broadly consistent with the experimental and numerical observations of chaotic vortex dynamics~\cite{Neely2012a}.

We emphasize that the quantum Reynolds number estimate proposed here is applicable to a finite-temperature weakly interacting superfluid and may provide a general condition in analogy with classical fluids that is independent of dimension. However, taking the zero-temperature limit gives an infinite value for $Re_q^0$, and in this regime the superfluid dissipation stems from vortex reconnections (or annihilation in two dimensions) and coupling to the sound field~\cite{Volovik2003a,Barenghi08a,PhysRevLett.97.145301}. The detailed description of criteria for superfluidity in the cross-over from the zero-temperature to high-temperature regimes is an open problem~\cite{Finne2003a}. Furthermore, due to the reversed direction of energy transfer in 2DQT, the scales of interest have to be reexamined; we do not pursue this here. Our aim is to establish a conceptual link between the dGPE and the NSE, given by Eq.~\eref{qnsdef}. In doing so we have shown how to identify the equivalent viscosity in a finite temperature BEC.

\subsection{Two-dimensional vortex wavefunction}
In the remainder of this work we limit our analysis to homogeneous compressible quantum fluids in two dimensions, and redefine our spatial and velocity coordinates accordingly: $\mathbf{r}=(x,y)=r\,(\cos\theta,\sin\theta)$ and $\mathbf{v}=(\mathrm{v}_x\,,\mathrm{v}_y)$. We thus confine our attention to the regime of an effective 2D GPE, with modified interaction parameter.
While 2D BEC systems can be created through extremely tight confinement in one dimension, a regime of effective 2D vortex dynamics can also be accessed in less oblate systems, giving a 2D analysis wider applicability. For example, although the BECs of References \cite{Neely10a} and \cite{Neely2012a} were three dimensional, the confinement along one dimension was strong enough to limit vortex motion to a plane and suppress vortex bending and tilting away from the tight-trapping direction.  Aspects of BEC dimensionality in regards to vortices and Kelvin waves were analyzed in \cite{Rooney11a}, further indicating that sufficiently oblate 3D BECs may be considered 2D as far as vortex dynamics and turbulence are concerned.  At the same time, such systems can remain far enough away from the quasi-2D limits in which a Berezhkinski-Kosterlitz-Thouless (BKT) transition has been observed \cite{Hadzibabic08a}, and BKT physics may thus be neglected. 

For our analysis of kinetic energy spectra, we require certain properties of a quantized vortex, namely the asymptotic character of the wavefunction for large and small length scales. The Gross-Pitaevskii equation describing the homogeneous ($V=0$) 2D Bose gas is obtained from \eref{eom} by taking $\gamma=0$ and using an interaction parameter $g_2 = g/l$ where $l$ is the characteristic thickness of the 3D system \cite{Numasato10b}:
\EQ{\label{GPEdef}
i\hbar\frac{\partial \psi(\brr,t)}{\partial t}&=&\left(-\frac{\hbar^2\nabla_\perp^2}{2m}+g_{2}|\psi(\brr,t)|^2\right)\psi(\brr,t).
}
For example, in a system with harmonic trapping in the $z$-direction characterized by trapping frequency $\omega_z$, the length scale is $l=\sqrt{2\pi}l_z$ where $l_z=\sqrt{\hbar/m\omega_z}$ is the $z$-axis harmonic oscillator length, and the confinement is assumed sufficient to put the wavefunction into the $z$-direction single-particle ground state. 

For solutions with chemical potential $\mu$ containing a single vortex at the origin (with circulation normal to the plane of the quantum fluid) we can write~\cite{Fetter2001}
\EQ{\label{oneVdef}
\psi_1(\brr,t)=\sqrt{n_0}e^{-i\mu t/\hbar}\chi\!\left(r/\xi\right)e^{\pm i\theta}
}
where $\xi=\hbar/m c$ is the healing length for speed of sound $c=\sqrt{\mu/m}$, and $n_0=\mu/g_2$ is the 2D particle density for $r\gg \xi$ and is taken to be a constant. The vortex radial amplitude function $\py(\pxc)$, where $\pxc = r/\xi$ is a scaled radial coordinate, is a solution of 
\EQ{\label{yeq}
\left(-\pxc^{-1}\partial_\pxc\, \pxc\partial_\pxc+\pxc^{-2}\right)\py=2(\py-\py^3).
}
The boundary conditions are $\py(0)=0$, and the derivative $\py^{\prime} \equiv d \py/d\pxc$ evaluated at $\pxc=0$ must be chosen such that it is consistent with $\py(\infty)=1$ and $\py^\prime(\infty)=0$. The value 
\EQ{\label{coreDeriv}
\Lambda\equiv \py^\prime(0)=\lim_{r\to 0}\frac{\xi}{\sqrt{n_0}}\left|\frac{d\psi_1}{dr}\right|
}
is determined numerically to be $\Lambda =0.8249\dots$. The state (\ref{oneVdef}) has the velocity field of a quantum vortex
\EQ{\label{vortV}
\mathbf{v}(\brr)=\frac{\hbar}{mr}(\mp\sin\theta,\pm\cos\theta).
}

\subsection{Kinetic energy decomposition}
We make use of the decomposition of the kinetic energy into compressible and incompressible parts~\cite{Nore97a,Horng09a}. 
The 2D case of the Gross-Pitaevskii energy functional \eref{hc} can be decomposed as $E=E_{K}+E_{V}+E_{I}+E_Q$, where
\EQ{\label{EtermsK}
E_K&=&\frac{m}{2}\int d^2\brr\; \rho(\brr,t)|\mathbf{v}(\brr,t)|^2,\\\label{EtermsV}
E_V&=&\int d^2\brr\; \rho(\brr,t)V(\brr,t),\\\label{EtermsI}
E_I&=&\frac{g_2}{2}\int d^2\brr\; \rho(\brr,t)^2,\\\label{EtermsQ}
E_Q&=&\frac{\hbar^2}{2m}\int d^2\brr\; |\nabla\!\sqrt{\rho(\brr,t)}|^2.
}

Respectively, these define the components of energy that can be attributed to kinetic energy, potential energy, interaction energy, and quantum pressure.  We are interested in the kinetic energy, $E_K$.   We define a density-weighted velocity field $\mathbf{u}(\brr,t)\!\equiv\!\!\sqrt{\rho(\brr,t)}\mathbf{v}(\brr,t)$, then  decompose this into $\mathbf{u}(\brr,t)=\mathbf{u}^i(\brr,t)+\mathbf{u}^c(\brr,t)$, where the incompressible field $\mathbf{u}^i$ satisfies $\nabla\cdot \mathbf{u}^i=0$, and the compressible field $\mathbf{u}^c$ satisfies $\nabla\times \mathbf{u}^c=0$.  We can further decompose the kinetic energy as $E_K=E_{i}+E_{c}$, where the portion of $E_K$ attributed to compressible or incompressible kinetic energy is defined as
\EQ{\label{Kedec}
E_{c,i}=\frac{m}{2}\int d^2\brr\; |\mathbf{u}^{c,i}(\brr,t)|^2.
}
The compressible component is attributed to the kinetic energy contained in the sound field, while the incompressible part gives the contribution from quantum vortices.  Our analysis below only involves $E_i$.  Because we focus on vortex configurations at instants in time, we drop the explicit time dependence from the remainder of our expressions.

In $k-$space, the total incompressible kinetic energy $E_i$ is  given by
\EQ{\label{ikek}
E_i=\frac{m}{2}\sum_{j=x,y}\int d^2\bk\;|{\cal F}_j(\bk)|^2
}
where
\EQ{\label{Fj}
{\cal F}_j(\bk)=\frac{1}{2\pi}\int d^2\brr\; e^{-i\bk\cdot \brr}\mathbf{u}^i_j(\brr).
}
The one-dimensional \emph{spectral density} in $k$-space is given in polar coordinates by integrating over the azimuthal angle to give
\EQ{\label{ikeks}
E_i(k)=\frac{mk}{2}\sum_{j=x,y}  \int_0^{2\pi}d\phi_k|{\cal F}_j(\bk)|^2,
}
which, when integrated over all $k$, gives the total incompressible kinetic energy $E_i=\int_0^\infty dk\; E_i(k)$. 

\section{Incompressible kinetic energy spectra}
\subsection{Incompressible kinetic energy spectrum of a vortex}\label{sec:onevort}
We now consider the kinetic energy spectrum of a single quantum vortex in a 2D BEC.
For an arbitrary wavefunction the decomposition into compressible and incompressible parts must be performed prior to carrying out the transformation to the spectral representation. However, for a quantum state containing a single vortex and no acoustic energy [i.e. the single vortex wavefunction $\psi_1$ \eref{oneVdef}] we note that the wavefunction is automatically incompressible, i.e. the compressible part is identically zero:
\EQ{\label{onev}
\nabla\cdot (\sqrt{\rho(\brr)}\mathbf{v}(\brr))=\mathbf{v}\cdot\nabla\sqrt{\rho(\brr)}+\sqrt{\rho(\brr)}\nabla\cdot \mathbf{v}\equiv 0.
}
The first term vanishes due to the orthogonality of the density gradient and velocity of a vortex, and the second due to the form of \eref{vortV}. Thus the incompressible spectrum is the entire spectrum for a single quantum vortex. 

\begin{figure}[!t]
\begin{center}
\includegraphics[width=\columnwidth]{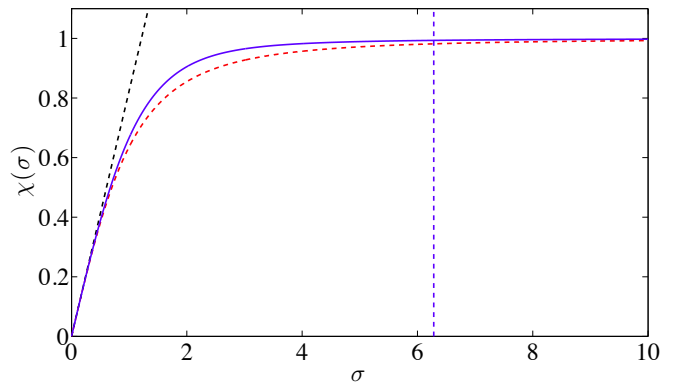}
\caption{Amplitude of the wavefunction for a single vortex solution of the Gross-Pitaevskii equation. The numerical solution of \eref{yeq} (solid line) is compared with the ansatz \eref{vort} (dashed line). The inclined dashed line shows the slope $\Lambda$ of the exact solution at the origin. The vertical line is the point $\sigma=2\pi$.  \label{figx}  }
\end{center}
\end{figure}

For a single vortex we can thus ignore the incompressible decomposition and cast the kinetic energy spectrum in terms of the properties of the radial amplitude function $\py(\pxc) = \sqrt{\rho(\pxc\xi)/n_0}$ obtained from \eref{yeq}. We have 
\EQ{\label{Fxfull}
{\cal F}_x(\bk)&=&-\frac{\hbar}{2\pi m}\int d^2\brr\; e^{-i\bk\cdot\brr}\frac{\sqrt{\rho(\brr)}}{r}\sin\theta\nonumber\\
&=&\frac{i\hbar}{m}\frac{d}{dk}\int_0^\infty dr\frac{\sqrt{\rho(r)}}{r}J_0(kr)\nonumber\\
&=&\frac{-i\hbar\sqrt{n_0}\xi}{m}\frac{1}{k\xi}\int_0^\infty d\pxc\;\py^\prime(\pxc)J_0(k\xi \pxc),
}
where $J_0$ is the zeroth-order Bessel function of the first kind.  Similar analysis gives ${\cal F}_y(k)=-{\cal F}_x(k)$. We can thus find the one-vortex spectrum [see \eref{ikeks}]
\EQ{\label{E1full}
E_i^1(k)=\Omega\xi^3F(k\xi),
}
where we define the dimensionless {\em core spectral function} 
\EQ{\label{Fc}
F(\pz)\equiv \frac{1}{\pz}\left(\int_0^\infty d\pxc\;\py^\prime(\pxc)J_0(\pz \pxc)\right)^2,
}
and we have introduced the unit of enstrophy
\EQ{\label{ensunit}
\Omega\equiv \frac{2\pi\hbar^2 n_0}{m\xi^2},
}
giving $\Omega\xi^3$ as the natural unit for the kinetic energy density. The core spectral function has the small-$\pz$ asymptotic form
\EQ{\label{Fcsmallz}
F(\pz)\Big|_{\pz\ll 1}=\frac{1}{\pz}\left(\int_0^\infty d\pxc\;\py^\prime(\pxc)\right)^2&=&\frac{1}{\pz}.
}
For $\pz\gg1$, $J_0(\pz\pxc)$ is highly oscillatory except at $\pxc=0$ where it is unity, and the Taylor expansion of $\py^\prime(\pxc)$ can be truncated at zeroth order to give
\EQ{\label{Fcbigz}
F(\pz)\Big|_{\pz\gg 1}&=&\frac{\Lambda^{2}}{\pz}\left(\int_0^\infty d\pxc\;J_0(\pz\pxc)\right)^2=\frac{\Lambda^{2}}{(\pz)^3}.
}
We thus have the asymptotic spectra for a single vortex
\EQ{\label{Ek1smallk}
E_i^1(k)\Big|_{k\xi\ll1}&=& \frac{\Omega\xi^3}{ k\xi}, \\
\label{Ek1largek}
E_i^1(k)\Big|_{k\xi\gg1}&=&\Lambda^{2}\frac{\Omega\xi^3}{(k\xi)^3}.
}
The $k\xi \ll1$ regime arises purely from the irrotational velocity field of a quantum vortex, while the $k\xi\gg 1$ regime is a property of the core of a compressible quantum vortex. The $k\xi \gg 1$ regime explicitly depends on the slope of the wavefunction at the core of a vortex. The cross-over between these regions occurs in the vicinity of $k\xi \approx 1$, hence we take $k\xi=1$ as distinguishing the infrared ($k\xi<1$) and ultraviolet ($k\xi>1$) regimes in the remainder of our analysis. The scale $k\xi=1$ thus serves to define an important length scale of the problem, namely $l_v\equiv 2\pi\xi$. In Fig.~\ref{figx} we see that at this distance from the vortex core the deviation of the amplitude from the background value is very small. This is the scale beyond which the details of the core structure are no longer important in characterizing the wavefunction, or equivalently that the fluid density has approximately reached its bulk value. The irrotational velocity field in Eq.~\eref{vortV} is the only remaining signature of a vortex at this range from its center and beyond. We note that our derivation of the $k^{-3}$ power-law stemming from the quantum vortex core structure is consistent with recent analysis of the Kelvin-wave cascade in 3D~\cite{Krstulovic2010a}.

Now that we have identified the properties of a single vortex, it is natural to ask whether a unit of enstrophy can be attributed to a single quantum vortex, and to compare this with the quantity defined in \eref{ensunit}. The point-vortex model suggests that this can be done, but gives a singular result, which is nevertheless known to be proportional to the number of vortices~\cite{Numasato10b}. The problem is also evident if we attempt to evaluate the enstrophy of a single vortex from the spectrum \eref{E1full}. Multiplying by $k^2$ to produce an enstrophy spectral density and integrating this over the ultraviolet regime $k\xi>1$, we are faced with the singular integral $\int_1^\infty dk/k$.

In this work we therefore define a new asymptotic quantity with units of enstrophy as
\EQ{\label{ons}
\zeta\equiv\lim_{k\to \infty}k^3E_{i}(k).
}
This quantity plays a fundamental role for a compressible superfluid because it completely specifies the large-$k$ region of the incompressible kinetic energy spectrum. Because the spectrum in this region of $k$-space is determined by the core structure of quantized vortices, we call this unit the {\em onstrophy} to both recall Onsager's contribution to our understanding of quantized vorticity in a superfluid and emphasize the difference between enstrophy in classical and quantum fluids. For a single quantum vortex we find, using \eref{E1full}, the onstrophy
\EQ{\label{ons1}
\zeta_1=\Lambda^{2}\Omega,
}
which differs from \eref{ensunit} by the factor $\Lambda^2=0.6805\dots$, a property of the vortex core in a compressible superfluid. 

\subsection{Vortex wavefunction ansatz and kinetic energy spectrum}
To study 2D kinetic energy spectra we will make extensive use of an algebraic ansatz for the wavefunction of a single vortex in a homogeneous superfluid. Numerical evaluation of the exact core function \eref{Fc} is not straightforward due to the highly oscillatory integrand, and the need to determine the vortex amplitude $\py(\pxc)$ extremely accurately over a large range of length scales. In order to accurately represent the spectrum it will be crucial that our ansatz have the correct asymptotic properties for small and large length scales described immediately above \eref{coreDeriv}. 
Making use of the slope at the origin computed for the exact solution in \eref{coreDeriv}, we use the ansatz wavefunction:
\begin{equation}\label{vort}
\phi_v(\brr)=\sqrt{n_0}\frac{re^{\pm i\theta}}{\sqrt{r^2+(\Lambda^{-1}\xi)^2}}.
\end{equation}
The general form of this ansatz has been previously used to describe the shape of a vortex core \cite{Fetter2001}, but here we use a length scale $\Lambda^{-1}\xi$ that enforces matching the slope of the ansatz density distribution to the exact value at the center of the core.  
The state \eref{vort} has the irrotational velocity field of a quantum vortex specified in \eref{vortV} and reproduces the asymptotic slope of the exact solution near the origin, as shown in Figure \ref{figx}. 

\begin{figure}[!t]
\begin{center}
\includegraphics[width=\columnwidth]{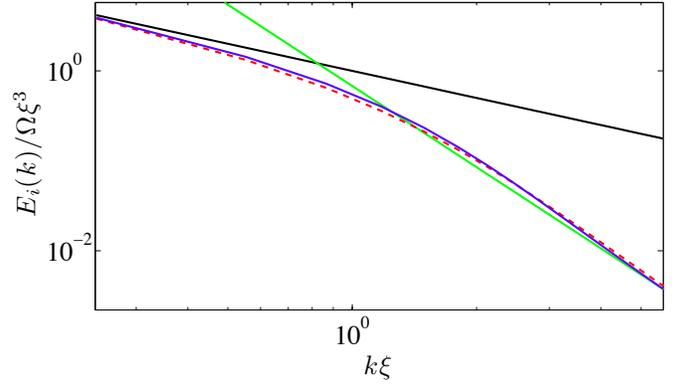}
\caption{Comparison of the numerically computed kinetic energy spectrum [\eref{E1full}, blue solid line] and that obtained from the core function $F_\Lambda(k\xi)$ [\eref{Ekv}, red dashed line] in the cross-over regime $k\xi\sim 1$. The asymptotic expressions \eref{Fcsmallz} and \eref{Fcbigz} are shown by the black and green lines respectively.
\label{coreFun}}
\end{center}
\end{figure}

We now compute the kinetic energy spectrum for a single vortex by evaluating \eref{Fj} using the form \eref{vort}. Taking $\Lambda^{-1}\xi=\vcore$ for brevity, we have
\EQ{\label{Fx}
{\cal F}_x(\bk)&=&\frac{i\hbar\sqrt{n_0}}{m}\frac{d}{dk}\int_0^\infty\frac{dr\;J_0(kr)}{\sqrt{r^2+\vcore^2}}\nonumber\\\label{Fxval}
&=&i\frac{\hbar\sqrt{n_0}\vcore}{2m}\left[I_1\left(\frac{k\vcore}{2}\right)K_0\left(\frac{k\vcore}{2}\right)-I_0\left(\frac{k\vcore}{2}\right)K_1\left(\frac{k\vcore}{2}\right)\right]\;\;\;\;
}
where $I_j$ and $K_j$ are modified Bessel functions of the first and second kind, respectively, of order $j$. Since $|{\cal F}_y|^2$=$|{\cal F}_x|^2$, we find the incompressible energy spectrum of a single vortex
\EQ{\label{Ekv}
E_i^1(k)=\Omega\xi^3F_\vslope (k\xi),
}
where 
\EQ{
F_\vslope(\pz)\equiv\Lambda^{-1} f(\pz\Lambda^{-1}),
}
and where we define
\EQ{\label{Fun}
f(z)\equiv (z/4)[I_1(z/2)K_0(z/2)-I_0(z/2)K_1(z/2)]^2.
}
The function $f(z)$ has the following asymptotics: for $z\ll 1$
\EQ{\label{zsmall}
f(z)=\frac{1}{z}+\left(\bar{\gamma}+\ln \left(\frac{z}{4}\right)\right)\frac{z}{2}+\dots,
}
where $\bar{\gamma}=0.57721...$ is the Euler-Masceroni constant;
for $z\gg$1
\EQ{\label{zlarge}
f(z)=\frac{1}{z^3}+ \frac{3}{ z^5}+\dots.
}
The function $F_\vslope(\pz)$ thus has the asymtotics
\begin{eqnarray}
F_\vslope(\pz)\Big{|}_{\pz\ll1}&=&\frac{1}{\pz}\\
F_\vslope(\pz)\Big{|}_{\pz\gg1}&=&\frac{\Lambda^2}{(\pz)^3}
\end{eqnarray}
which are identical to those of $F(\pz)$. The two functions are very similar, with only small differences evident in the cross-over region $\pz\sim 1$, as seen in Figure \ref{coreFun}. We use $F_\Lambda(\pz)$ instead of $F(\pz)$ for describing the kinetic energy spectrum for a vortex core in the remainder of this work as it is numerically expedient and does not alter any of the physical consequences of our analysis. Towards the end of this paper we will compare the asymptotic results of our analysis with spectra determined numerically from forced dGPE dynamics.

The spectrum of a single vortex is shown in Figure \ref{fig1}, and compared with the spectrum of a vortex-antivortex pair (a vortex dipole), and that for two vortices of the same sign (a vortex pair).   These two-vortex spectra are analyzed in the following section.

\subsection{Two-vortex spectra}
Extension of the discussion in Section \ref{sec:onevort} leads us to conclude that a wavefunction that \emph{only} contains vortices (i.e., no sound field) separated by more than a few healing lengths will thus be approximately incompressible according to the decomposition. The approximation breaks down through the non-orthogonality of $\mathbf{v}$ and $\nabla\sqrt{\rho(\brr)}$ near a vortex core due to the velocity field induced by the other vortices. However, in the close vicinity of a vortex core, where $\nabla\sqrt{\rho(\brr)}$ is significant, the velocity is dominated by the velocity field of that vortex core. An arrangement of vortices separated by more than a few healing lengths will thus be approximately incompressible. In the following analytical treatment we will neglect any compressible part that arises from an assembly of vortices described by the ansatz \eref{vort}.

\begin{figure}[!t]
\begin{center}
\includegraphics[width=\columnwidth]{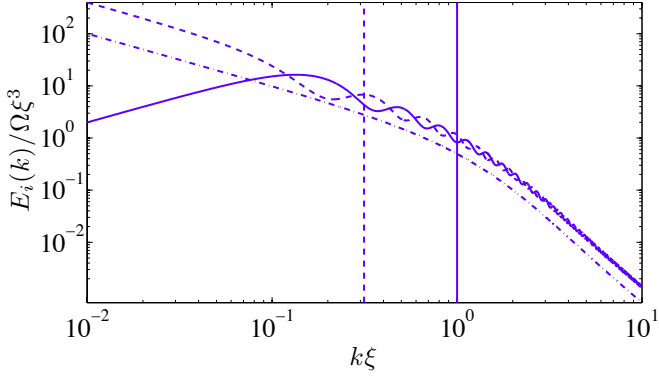}
\caption{Incompressible kinetic energy spectra for a single vortex (chain line), a vortex dipole (solid line), and a vortex pair (dashed line). The vortex dipole and pair are both shown for vortex separation $d=20\xi$, and the wavenumber $k_d\equiv 2\pi/d$ is shown as a vertical dashed line. The cross-over scale $k\xi=1$ is given by the solid vertical line.
\label{fig1}}
\end{center}
\end{figure}
A two-vortex state in a homogeneous system with no boundaries has density-weighted velocity field 
$\sqrt{\rho(\brr)}[\mathbf{v}_1(\brr)+\mathbf{v}_2(\brr)]$, where $\mathbf{v}_j$ is the velocity field around vortex $j = 1,2$ taken separately. If the vortex cores are separated by $d\gg \xi$ it will also be a very good approximation to write $\rho(\brr)=\rho_1(\brr)\rho_2(\brr)/n_0$, with 
\begin{equation}\label{twovort}
\rho_j(\brr)=\frac{n_0\;|\brr - \brr_j|^2}{|\brr - \brr_j|^2+(\Lambda\xi^{-1})^2}.
\end{equation}
The density weighted two-vortex velocity field can then be written as
\EQ{\label{wapprx}
\sqrt{\frac{\rho_1(\brr)\rho_2(\brr)}{n_0}}[\mathbf{v}_1(\brr)+\mathbf{v}_2(\brr)]&=&\sqrt{\rho_1(\brr)}\mathbf{v}_1(\brr)+\sqrt{\rho_2(\brr)}\mathbf{v}_2(\brr)\nonumber\\
&&+K_{12}(\brr).
}
The final term is
\EQ{\label{Kerr}
K_{12}(\brr)&=&
\sqrt{\frac{\rho_1(\brr)\rho_2(\brr)}{n_0}}\Bigg(\mathbf{v}_1(\brr)\left[1-\sqrt{\frac{n_0}{\rho_2(\brr)}}\right]\nonumber\\
&&+\mathbf{v}_2(\brr)\left[1-\sqrt{\frac{n_0}{\rho_1(\brr)}}\right]\Bigg)
}
which is only significant when considering the velocity field of one vortex in the close vicinity of the other vortex core. $K_{12}$ is therefore a negligible correction to the spectrum for widely separated vortices. This approximation may be trivially generalized to arbitrary numbers of vortices provided the cores do not overlap appreciably. This approximation is central to our treatment, as it allows recovery of familiar point-like vortex physics in the infrared regime $k\xi\ll 1$. Strict validity is limited to the regime where the intervortex spacing $d$ is bounded below by  $\sim l_v= 2\pi\xi$, the scale at which the core structure becomes evident (see Fig.~\ref{figx}).
\par
We require the transform
\EQ{\label{Fj2}
{\cal F}_j^d(\bk)&=&\frac{1}{2\pi}\int d^2\brr\; e^{-i\bk\cdot \brr}\left[\sqrt{\rho_1(\brr)}\mathrm{\textbf{v}}_1(\brr)\right]_j^i\nonumber\\
&&+\frac{1}{2\pi}\int d^2\brr\; e^{-i\bk\cdot \brr}\left[\sqrt{\rho_2(\brr)}\mathrm{\textbf{v}}_2(\brr)\right]_j^i
}
where the superscript $d$ denotes the case of a vortex dipole, and subscript $j = x,y$ indicates the $x$ and $y$ components of the density-weighted velocity fields of each vortex.  As above, the subscripts 1 and 2 denote vortices 1 and 2, and the superscript $i$ denotes that it is the incompressible, or divergence-free portion of the density-weighted field that is of interest here. To account for the opposite signs of circulation for the two vortices, without loss of generality we choose vortex 1 as positively charged, located at $\brr_0=(d/2)\hat{\x}$, so that $\mathbf{v}_1(\brr)=\mathbf{v}(\brr-\brr_0)$ where $\mathbf{v}(\brr)$ is the central vortex velocity field \eref{vortV}. Vortex 2 has velocity field $\mathbf{v}_2(\brr)=-\mathbf{v}(\brr+\brr_0)$. We then have
\EQ{\label{Fxdip}
{\cal F}_j^d(\bk)&=&\frac{1}{2\pi}\int d^2\brr\; e^{-i\bk\cdot (\brr+\brr_0)}\left[\sqrt{\rho(\brr)}\mathrm{\textbf{v}}(\brr)\right]_j^i\nonumber\\
&&-\frac{1}{2\pi}\int d^2\brr\; e^{-i\bk\cdot (\brr-\brr_0)}\left[\sqrt{\rho(\brr)}\mathrm{\textbf{v}}(\brr)\right]_j^i.
}
For a vortex dipole we can then write
\EQ{\label{Fdip}
{\cal F}_j^d(\bk)&=&{\cal F}_j(\bk) e^{-i\bk\cdot\brr_0}-{\cal F}_j(\bk) e^{i\bk\cdot\brr_0}
}
where ${\cal F}$ is the spectrum of a single vortex. Using \eref{Fxval}, and the fact that ${\cal F}_y(k)=-{\cal F}_x(k)$, we find for a vortex dipole
\EQ{\label{dipE}
E_i^d(k)=2\Omega\xi^3F_\Lambda(k\xi)(1-J_0(k d)).
}
The spectrum of a pair of vortices of the same circulation separated by $d$ is calculated similarly to be
\EQ{\label{vpair}
E_i^p(k)=2\Omega\xi^3 F_\Lambda(k\xi)(1+J_0(k d)).
}
The spectra \eref{dipE} and \eref{vpair} are shown in Figure \ref{fig1}. It is clear that for scales less than the vortex separation distance $d$ there is interference in $k$-space, leading to oscillations in the spectrum. The difference between the dipole and pair is that the interference fringes are offset, and the infrared asymptotics are different, a feature we discuss further below.  The spectrum of the vortex pair is clearly similar to that of the single vortex in the far infrared region, but the additional kinetic energy of the vortex pair state is observed throughout the spectrum.
\subsection{Kinetic energy spectrum of $N$-vortex configurations}
Extending the above analysis, for a general system of $N$ singly quantized vortices with circulation signs $\kappa_p=\pm 1$ located at $\brr_p$ the kinetic energy spectrum is given by
\EQ{\label{Nv}
E_{i}^N(k)=\Omega\xi^3 F_\Lambda(k\xi)\sum_{p=1,q=1}^N\kappa_p\kappa_q J_0(k|\brr_p-\brr_q |).
}
We note the resemblance to point-vortex results which also have the Bessel function dependence~\cite{Novikov1975,Numasato10a}. The function $F_\Lambda(k\xi)$ gives the incompressible limit for small $k$ [$J_0(kr)$ approaches unity for small $k$ and finite $r$], and introduces the physics of compressible superfluids for $1\lesssim k\xi$.
\par 
We can write (\ref{Nv}) as
\EQ{\label{NvE}
E_{i}^N(k)=N\Omega\xi^3F_\Lambda(k\xi)G_N(k)
}
where
\EQ{\label{Conf}
G_N(k)\equiv 1+\frac{2}{N}\sum_{p=1}^{N-1}\sum_{q=p+1}^N\kappa_p\kappa_q J_0(k|\brr_p-\brr_q|)
}
is a purely configurational function involving summation of $M=N(N-1)/2$ distinct intervortex distances. This function has the limits
\EQ{\label{CNlim1}
\lim_{k\to\infty}G_N(k)&=&1,\\\label{CNlim2}
\lim_{k\to 0}G_N(k)&=&1+\frac{2}{N}\sum_{p=1}^{N-1}\sum_{q=p+1}^N\kappa_p\kappa_q =\frac{\Gamma^2}{N},
}
where the total dimensionless circulation is defined by
\EQ{\label{circ}
\Gamma\equiv\frac{m}{\hbar}\oint_C \mathbf{v}\cdot d\mathbf{l}=\sum_{p=1}^N\kappa_p =N_+-N_-
}
for any contour $C$ enclosing all $N_+$ positive and $N_-=N-N_+$ negative vortices.

We then find that the onstrophy for the $N$-vortex system is
\EQ{\label{NVlim}
\zeta_N=\lim_{k\to \infty}\;k^3E_{i}^N(k)&=&N\Lambda^{2}\Omega=N\zeta_1,
}
and consequently
\EQ{\label{E_3}
E_{i}^N(k)\Big{|}_{k\xi\gg1}&=&\frac{\zeta_N}{k^3}=\frac{N\Lambda^2\Omega\xi^3}{(k\xi)^3}.
}
This is one of our central results: the ultraviolet regime $k\xi\gg\,$1 has a universal asymptotic form that is independent of the vortex configuration, and that resembles the ultraviolet spectrum of classical 2D turbulence that is identified with a direct cascade of enstrophy. If we try to evaluate the classical definition of enstrophy, the result is singular, yet the onstrophy definition \eref{ons} gives a well-defined additive quantity that is singularity free and depends only on the total number of vortices in the system.
\subsection{Infrared behavior}
When $\Gamma\neq 0$ the far infrared limit \eref{CNlim2} gives
\EQ{\label{kzero}
E_{i}^N(k)\Big{|}_{k\xi\ll1}&=&\frac{\Omega\xi^3\Gamma^2}{k\xi}.
}
This configuration-independent $k^{-1}$ power law arises from the far-field velocity distribution of a collection of point vortices, which becomes equivalent to that of a single vortex of charge $\Gamma$ at sufficiently large scales.

When $\Gamma\equiv 0$ we use the small-argument expansion $J_0(z)\simeq 1-z^2/4$,
and the asymptotic form \eref{zsmall}, to find the $k\xi\ll 1$ behavior determined by the configurational information contained in the intervortex distances $|\brr_p-\brr_q|$:
\EQ{\label{Elowk}
E_{i}^N(k)\Big{|}_{k\xi\ll1}=-\frac{\Omega\xi^2k}{2}\sum_{p=1}^{N-1}\sum_{q=p+1}^N\kappa_p\kappa_q |\brr_p-\brr_q|^2.
}
The simplest case involves a single vortex dipole and has only one length scale, namely the vortex separation, and the low-$k$ form $E_{i}^d(k)\simeq \Omega\xi^2d^2 k/2$, as shown in Fig.~\ref{fig1}. In general, when $\Gamma=0$ the infrared region of the spectrum is sensitive to the vortex configuration, but approaches a power-law for low-$k$ that has a configuration-independent exponent. The linear decay of kinetic energy as $k\to 0$ stems from the cancellation of the far-field velocity profiles for length scales greatly exceeding the largest intervortex separation in any neutral configuration of vortices.

\section{Kolmogorov spectrum}\label{scaleInvariantPlaw}
In the previous section we obtained an explicit expression \eref{NvE} for the incompressible kinetic energy spectrum that incorporates the compressible nature of individual vortex cores through the function $F_\Lambda(k\xi)$ (derived via an ansatz for the vortex core profile), which captures the essential physics of the corresponding exact solution  $F(k\xi)$ defined in \eref{Fc}.  For both functions, point-vortex physics is recovered at large length-scales ($k\xi\ll1$).
If the dynamical evolution is such that an inertial range associated with an inverse energy cascade develops, we should expect a Kolmogorov power law $E_i(k)\propto k^{-5/3}$ over the inertial range. It is clear from the form of \eref{NvE} that this law can only depend on the spatial configuration of the vortices. We now seek to understand the simplest situations that may show evidence for the existence of such an inertial range.  We consider forcing occurring via vortex and energy injection at a forcing scale $k_F\sim \xi^{-1}$, and describe vortex configurations that do and do not lead to a Kolmogorov law for $k<\xi^{-1}$.

We now assume an idealized case in which the infrared spectrum is continuous with the universal $k^{-3}$ law of the ultraviolet spectrum at the scale $k\xi\approx1$. This constraint imposes a strong restriction on the infrared spectrum, completely determining its form in the case that it satisfies a power law. In this respect the universal ultraviolet region has significant physical consequences. The power-law approximation to the universal ultraviolet region based on \eref{NVlim} has the form
\EQ{\label{uv}
E_{i,{\rm U}}^N(k)=\Lambda^{2}\frac{N\Omega\xi^3}{(k\xi)^3}=\zeta_Nk^{-3}.
}
The number of vortices determines the $N$-vortex onstrophy $\zeta_N$ \eref{NVlim}, from which the power-law approximation to the ultraviolet energy spectrum is completely determined. This power law is a very good approximation, as will be seen by sampling different vortex configurations below. The infra-red or configurational regime is then given by the $k\xi\ll 1$ regime of $F_\Lambda(k\xi)$:
\EQ{\label{irlim}
E_{i,\rm{C}}^N(k)\Big{|}_{k\xi\ll1}=\frac{N\Omega\xi^3}{k\xi}G_N(k).
} 
At this point we consider the consequences of assuming that a turbulent system will have a $k^{-5/3}$ law in the configurational regime, and that this power law is continuous with \eref{uv} at $k\xi = 1$. We suppose that $E_{i,{\rm C}}^N(k)\propto k^{-5/3}$. Continuity at $k=1/\xi$ then requires $E_{i,{\rm C}}^N(1/\xi)=E_{i,{\rm U}}^N(1/\xi)$, and gives the infrared spectrum
\EQ{\label{ir}
E_{i,{\rm C}}^N(k)=\Lambda^{2}\frac{N\Omega\xi^{3}}{(k\xi)^{5/3}}=\zeta_N \,\xi^{4/3} \, k^{-5/3}.
}
Thus the constraint that the universal regime is continuous at the cross-over scale $k=\xi^{-1}$ completely constrains the form of the configurational spectrum.  Physically this may correspond to an inertial range that extends upwards from the smallest scale of the infrared region given forcing at a wavenumber $k_F \sim \xi^{-1}$.  Note that this expression \eref{ir} has no reference to the signs of the vortex circulations,  the degree of circulation polarization, or vortex clustering.  By assuming continuity at $k\xi=1$ with a UV spectrum that has a universal $N$-vortex form, we have implicitly assumed that all $N$ vortices are involved in determining the spectrum of the inertial range. 

We might expect that this will give a very good description for a completely polarized system exhibiting fully developed turbulence. When there is clustering in mixtures of different sign vortices, the spectrum may well still approach a Kolmogorov law, but there is no reason to expect that it will cross over so smoothly. We return to this problem when we compare our analysis with numerical simulations of forced turbulence in Sec.~\ref{gpedyn}. 

It is useful at this point to give a simplified reiteration of Novikov's argument for the power law for the vortex distribution being $-1/3$ in Kolmogorov turbulence~\cite{Novikov1975}.
To obtain power-law behavior we must consider the spectrum for a vortex distribution involving many length scales. 
For simplicity we assume all vortices have the same sign of circulation, $\kappa_p\equiv\kappa$. The configuration function \eref{Conf} has $M=N(N-1)/2$ terms in the summation, and can be written as
\EQ{\label{configIR}
G_N(k)=1+\frac{2}{N}\sum_{p=1}^M J_0(k s_p),
}
in terms of an average over distinct vortex separations $s_i$. We introduce the intervortex distance distribution $P(s)$ such that $P(s)ds$ is the fraction of intervortex distances in the range $[s,s+ds)$. In the continuum limit  
\EQ{
G_N(k)&\propto& \int P(s)J_0(ks)ds.
}
We seek a distribution $P(s)$ that will generate a Kolmogorov law from the $N$-vortex spectrum \eref{NvE} for scales larger than the vortex core, $k\ll \xi^{-1}$, in the large-$N$ regime.
We then find from \eref{irlim} that
\EQ{\label{smallk}
E_{i, \rm{C}}^N(k)\Big{|}_{k\xi\ll1}&\sim&\frac{1}{k}\int_{\xi}^{\infty} P(s)J_0(ks)ds\\
&\simeq&\frac{1}{k}\int_{0}^{\infty} P(s)J_0(ks)ds.
}
The scale invariance of turbulence naturally leads to the assumption that the intervortex separation distribution is a power-law $P(s)\sim s^{-\alpha}$ over the scale range of interest. The requirement of a power law in the kinetic energy then gives the scaling relation
\EQ{\label{plaw}
E_{i,\rm{C}}^N(k) &\sim&\frac{1}{k^{\beta}}\sim \frac{1}{k}\int_{0}^{\infty} s^{-\alpha}J_0(ks)ds\nonumber\\
&=&\frac{1}{k^{2-\alpha}}\int_{0}^{\infty} \tau^{-\alpha}J_0(\tau)d\tau.
}
The integral is convergent for $-1/2<\alpha<1$, allowing 
\EQ{\label{betaRange}
1<\beta<5/2.
}
In particular, the universal Kolmogorov law $\beta=5/3$ occurs for
\EQ{\label{Pdist}
P(s)\sim s^{-1/3}
}
as obtained by Novikov~\cite{Novikov1975} for point-vortices.
We will test this scaling argument for the exponent in numerical sampling of {\em localized} vortex configurations in the following sections.
Testing if this vortex separation power-law holds in simulations and experiments may give a quantitative measure of fully developed 2D turbulence in a  compressible superfluid, and a way to identify the inertial range as the scale range over which this power-law can be identified. 

In 2D classical turbulence a $k^{-3}$ region of the kinetic energy spectrum is often associated with a direct enstrophy cascade. We note that this exponent $\beta=3$ is ruled out by \eref{betaRange}. Hence, within this continuum analysis the $k^{-3}$ power-law spectrum cannot occur in the configurational region for 2D quantum turbulence, as long as the vortex distribution follows a simple power law. This result suggests that if a direct enstrophy cascade were to occur in the configurational region of the spectrum, a different type of vortex distribution would be necessary.  This makes intuitive sense, since direct enstrophy cascades may be associated with the stretching of patches of vorticity in the 2D plane.
Furthermore, direct enstrophy cascades and energy spectra proportional to $k^{-3}$ have been noted in simulations of superfluid helium thin films~\cite{Chu2001a,Chu2001b}. Nevertheless, since we have also shown that the $k\xi>1$ range is determined entirely by the core structure, this gives a strong indication that a direct enstrophy cascade cannot occur in this region in compressible 2DQT.  

\par It is clear that configurations containing one or a few characteristic length scales, such as a vortex dipole or a vortex lattice, cannot lead to a power law spectrum for $E_{i,\rm{C}}^N(k)$. In the case of a vortex lattice the intervortex distance distribution has many discrete peaks~\cite{Bradley08a}. The vortex dipole and a vortex pair each have a single length scale and this leads to characteristic interference fringes in the energy spectrum seen in Fig.~\ref{fig1}. 

\subsection{Sampling spatial vortex distributions}\label{sec:vdist}
We now test our analysis of the spectrum by numerically sampling several vortex distributions $\{\mathbf{r}_p,\kappa_p\}_{p=1}^N$ and evaluating \eref{NvE}. A straightforward test of the statistical argument for the Kolmogorov power law to occur involves sampling the power law \eref{Pdist} and evaluating \eref{configIR}. Indeed, it is easily verified that this generates a $k^{-5/3}$ spectrum in the configurational region. However, the connection of such a sampling to particular spatial vortex distributions is not clear, and in fact the mapping is not unique. To make this connection concrete we require a way of sampling finite, localized, spatial vortex distributions for which the vortex separations are power-law distributed. 

In an ideal, infinitely extended vortex configuration exhibiting the power law \eref{Pdist}, the system is translationally invariant and the coordinate origin can be placed at any particular vortex, yielding the same power law for the radial distribution of vortices from the origin. In a finite system the scale invariance can only persist up to scales of order the largest vortex separation. Furthermore, the vortices must be separated by a minimum distance to satisfy the assumptions used in deriving the energy spectrum from the point-vortex model. In practice it is necessary to use a self-consistent sampling scheme in order to generate the correct power-law distributions for localized finite configurations. 

Our sampling scheme for a configuration of $N$ vortices is:
\begin{enumerate}
\item Sample the radial distance $r_p$ of each vortex from the coordinate origin according to a power law probability distribution $\propto r_p^{-\bar{\alpha}}$. The exponent $\bar{\alpha}$ is distinct from $\alpha$ due to the finite system size and localization of the distribution. 
\item Assign each vortex a randomly chosen, uniformly distributed angle $\theta_p\in[0,2\pi)$. The cartesian coordinates for vortex $p$ are then 
\EQ{\label{xy}
(x_p,y_p)&=&r_p(\cos\theta_p,\sin\theta_p).
}
\item For a given $N$, compute the kinetic energy spectrum, averaging over $n_s=100$ samples of vortex position data found via the foregoing routine. Iterate until the spectral power law of interest is found. 
\end{enumerate}

We will sample a power-law distribution of vortex distances from the origin. In practice there is a lower ($r_{\rm min}\sim \xi$) and upper ($r_{\rm min}\sim R=$ system size) cutoff for power-law scaling. We thus wish to sample the distribution
\EQ{\label{Pdistactual}
P_{\bar{\alpha}}(r)=\frac{1-\bar{\alpha}}{r_{\rm max}^{1-\bar{\alpha}}-r_{\rm min}^{1-\bar{\alpha}}}r^{-\bar{\alpha}},
}
that is normalized on the interval  $r_{\rm min}\leq r<r_{\rm max}$.
We sample $r$ values from this distribution using uniform random variates $x\in [0,1)$ via the transformation~\cite{Cla2009.SR51.661}:
\EQ{\label{rtrans}
r=\left[r_{\rm max}^{1-\bar{\alpha}}x+r_{\rm min}^{1-\bar{\alpha}}(1-x)\right]^{\frac{1}{1-\bar{\alpha}}}.
}
Note that when $\bar\alpha<1$, as is always the case in this work, some kind of ultraviolet cutoff is required for the distribution to be normalizable. Here we have made a choice that gives a power-law distribution over a well-defined scale range (See e.g. Ref.~\cite{Cla2009.SR51.661} for other common choices). 

\begin{figure*}[!t]
\begin{center}
\includegraphics[width=\textwidth]{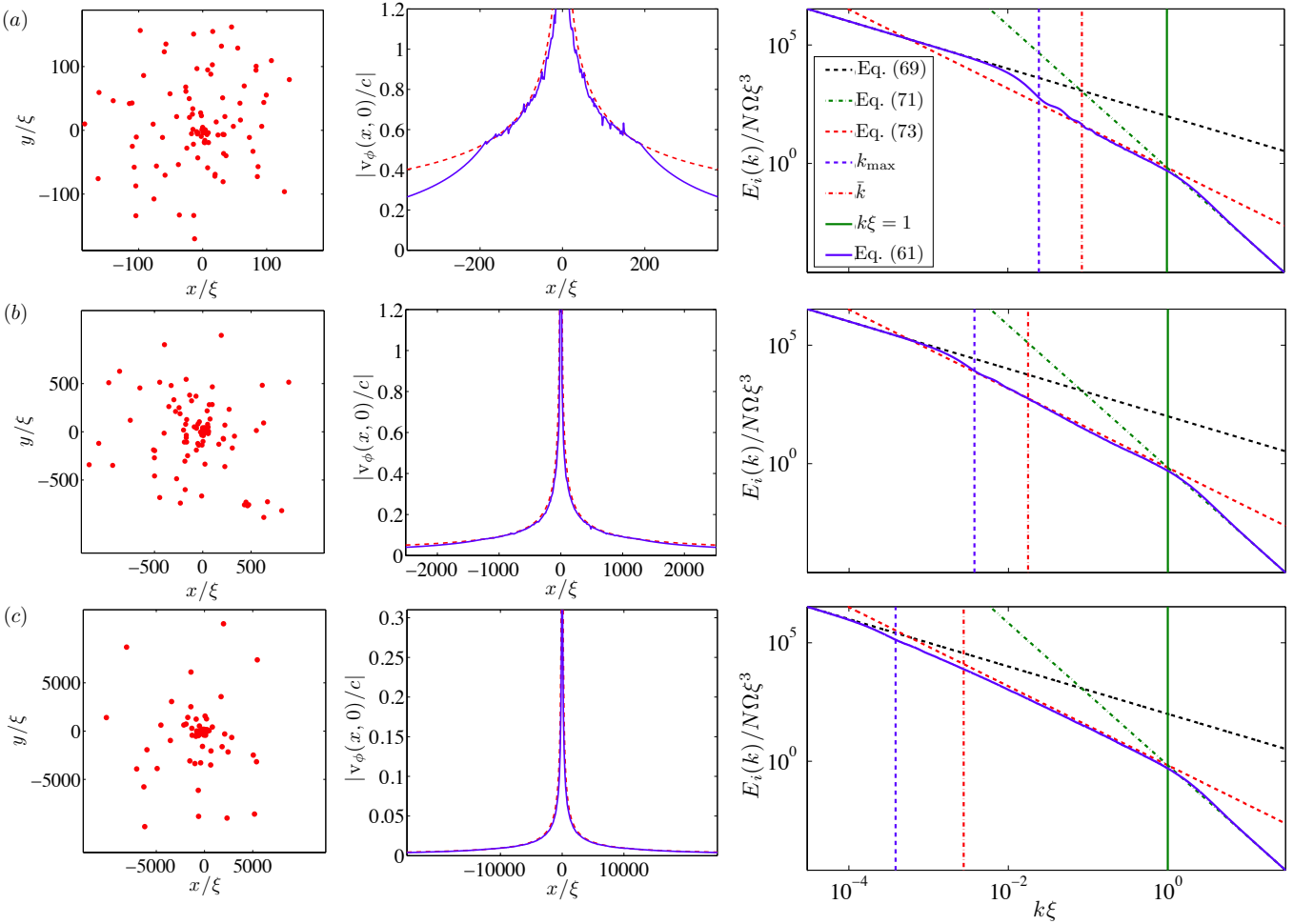}
\caption{Locations of $N=100$ vortices for particular samples (left), with velocity distributions [middle, with the analytical result \eref{vprof} (dashed line)], and kinetic energy spectra (right, blue curves), averaged over $n_s=100$ samples distributed according to \eref{xy} and \eref{Pdistactual}, with $r_{\rm min}=\xi$, and (a) $r_{\rm max}=60\pi\xi$, $\bar\alpha=0.4$, (b) $r_{\rm max}= 400\pi\xi$, $\bar\alpha=0.65$, and (c) $r_{\rm max}=4000\pi\xi$, $\bar\alpha=0.83$. The increasing scale range of power-law behavior gives an inertial range in the corresponding kinetic energy spectrum of $\sim 1$, 1.5, and 2 decades respectively. \label{randAll}}
\end{center}
\end{figure*}
 \subsection{Classical velocity distribution of a large cluster}
A fully polarized configuration of vortices with a given radial power law forms a quantum analogue of the coherent vortices of forced 2D turbulence in classical fluids. As an arrangement of many vortices, the velocity distribution must approach a classical limit, according to Bohr's correspondence principle, in much the same way that a rotating Abrikosov lattice generates a velocity field that approaches that of a rotating rigid body~\cite{Fetter2001}. In what follows we find the classical velocity field, and identify the physical significance of the radial exponent $\bar{\alpha}$.

To determine the velocity field we first compute the fraction of vortices enclosed by a circular contour around the origin, with radius $r$, for the distribution \eref{Pdistactual}. Taking $r_{\rm min}=\xi, r_{\rm max}=R$, and considering scales $\xi\ll r\ll R$ we have
\EQ{\label{fracV}
f_r&=&\int_{\xi}^r P_{\bar{\alpha}}(u)du={\cal N}\int_{\xi}^r \frac{du}{u^{\bar{\alpha}}}=\frac{{\cal N} }{1-\bar{\alpha}}\left(r^{1-\bar{\alpha}}-\xi^{1-\bar{\alpha}}\right)\nonumber\\
&\simeq&\left(\frac{r}{R}\right)^{1-\bar{\alpha}},
}
where we used the fact that ${\cal N}^{-1}=\int_{\xi}^R s^{-\bar{\alpha}}ds=(1-\bar{\alpha})^{-1}(R^{1-\bar{\alpha}}-\xi^{1-\bar{\alpha}})\simeq R^{1-\bar{\alpha}}/(1-\bar{\alpha})$ normalizes the distribution up to the largest scale $R$. 
Considering the average azimuthal velocity component $ \mathrm{v}_\phi(r)$, the circulation is
\EQ{\label{circ}
\oint \mathbf{v}\cdot d\mathbf{l}=\frac{h}{m}n= \mathrm{v}_\phi(r)2\pi r
}
where $n=f_rN$ is the number of vortices enclosed by the contour of radius $r$. Using \eref{fracV}, we obtain the velocity profile
\EQ{\label{vprof}
\mathrm{v}_\phi(r)\simeq\frac{cN}{(R/\xi)^{1-\bar{\alpha}}(r/\xi)^{\bar{\alpha}}}.
}
This {\em inertial cluster} has a power-law velocity profile determined by the specific radial exponent $\bar{\alpha}$. In contrast, the velocity profile of an Abrikosov vortex lattice rotating at frequency $\omega$ approaches that of a rigid body $ \mathrm{v}_\phi(r)=\omega r$, and a single quantum vortex has profile $ \mathrm{v}_\phi(r)=\hbar/mr$. The inertial cluster velocity profile \eref{vprof} is compared with sampled distributions (see below) in Figure \ref{randAll}.
\subsection{Scale expansion of a large cluster}
We now illustrate the role of the radial power law exponent $\bar\alpha$ and the classical velocity distribution by sampling a large cluster. We vary our choice of scale range for the vortices ($r_{\rm max}$ in Eq.~\ref{Pdistactual}), and investigate how the range of $k^{-5/3}$ changes.

A characteristic wave number measuring cluster size for a sample involving $n_c$ vortices in a given cluster is given by
\EQ{\label{kbar}
\bar{k}\equiv 2\pi/\bar{r}
}
Where $\bar{r}=\frac{1}{n_c}\sum_{p=1}^{n_c}r_p$. In the figure we plot the scale $\bar{k}$ to give an indication of the range of $k^{-5/3}$ scaling. We also plot $k_{\rm max}$ corresponding to the largest vortex separation scale in the system
\EQ{\label{maxr}
k_{\rm max}=2\pi/{\rm max}|{\bf r}_p-{\bf r}_q|.
}
For $k<k_{\rm max}$ the velocity field approaches that of a single vortex.
\par
In Figure \ref{randAll} vortex distributions are sampled for $N=100$ vortices of the same sign using the sampling scheme \eref{xy}, \eref{rtrans}, for $r_{\rm min}=\xi$ and different values of $r_{\rm max}$. Individual samples are shown to indicate the spread of vortices, and the velocity profiles and kinetic energy spectra are computed by averaging over $n_s=100$ samples. The mean azimuthal velocity compares well with Equation \eref{vprof}, showing that the specific radial exponent $\bar\alpha$ in \eref{Pdistactual} also determines the power law of the azimuthal velocity field, as seen in \eref{vprof}. For scales larger than $r_{\rm max}$, $v_\phi(r)$ returns to the $r^{-1}$ scaling for a charge $N$ vortex [clearly seen in Fig.~\ref{randAll}\;$(a)$]. $N=100$ vortices distributed up to $r_{\rm max}=60\pi\xi$ gives an inertial range in the kinetic energy spectrum of approximately one decade, for $\bar\alpha=0.4$ (as the number of vortices in a given scale range increases, $\bar \alpha\to 1/3$). Increasing the upper scale cutoff to $r_{\rm max}=400\pi\xi, 4000\pi\xi$ gives $\bar\alpha=0.65,0.83$, with 1.5 and 2 decades of inertial range respectively. Thus expanding the scale range of power-law behavior expands the inertial range, but requires the azimuthal velocity profile to steepen. In the UV region of the spectrum the $k^{-3}$ law always holds [Eq.~\eref{uv}], while the inertial range holds for $ \bar k\lesssim k\lesssim \xi^{-1}$, and single vortex behavior is apparent for $k< k_{\rm max}$. 
\begin{figure}[!t]
\begin{center}
\includegraphics[width=\columnwidth]{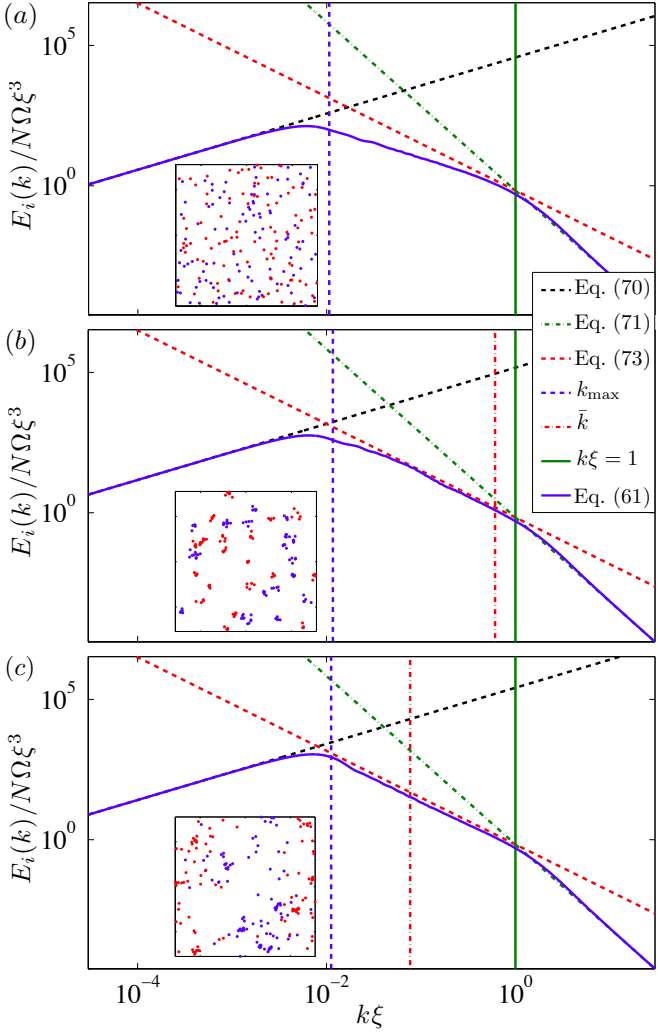}
\caption{Kinetic energy spectra for $N_+=100$, and $N_-=100$ vortices, averaged over $n_s=100$ samples (blue curves). The vortices are distributed (a) uniformly  over the $(600\xi)^2$ square domain (inset), and clusters are sampled according to \eref{xy} and \eref{Pdistactual}, with (b) $n_c=5$ vortices in each cluster, $r_{\rm min}=\xi$, $r_{\rm max}=10\pi\xi$, and (c) $n_c=20$ vortices in each cluster and $r_{\rm max}= 100\pi\xi$, with each cluster center uniformly distributed over the periodic domain as in (a). The power-law sampling in figures (b) and (c) requires a radial exponent of $\bar\alpha= 0.8$, and $0.75$, and gives a Kolmogorov $k^{-5/3}$ region in the corresponding kinetic energy spectrum of $\sim 1$ and 2 decades respectively.\label{100clustering}}
\end{center}
\end{figure}
\subsection{Clustering in a neutral distribution}
We now consider the role of increasing clustering in expanding the inertial range of the kinetic energy spectrum. First, in $n_s=100$ samples, we distribute $N_+=N_-=100$ vortices over a uniform periodic domain; one such sample is shown in the inset of Figure \ref{100clustering} (a). The corresponding kinetic energy spectrum shows the correct UV-region spectrum given by Eq.~\eref{uv}, and also approaches the $E_i(k)\propto k$ form for $k<k_{\rm max}$, given by Eq.~\eref{Elowk}. For $k_{\rm max}\lesssim k\lesssim \xi^{-1}$ the spectrum is less steep than $k^{-5/3}$ and the system lacks an inertial range. In Figure \ref{100clustering} (b) the vortices are sampled as 40 clusters of $n_c=5$ vortices of the same sign according to \eref{xy} and \eref{rtrans}, with $\bar\alpha=0.8$, and $r_{\rm min}=\xi$, $r_{\rm max}=10\pi\xi$. The 10 +ve and 10 -ve cluster centers are uniformly distributed as in Fig.~\ref{100clustering} (a), as seen in the sample (inset). This distribution yields a $k^{-5/3}$ power-law kinetic energy spectrum over  $\sim1$ decade of wave numbers. By further expanding the scale of clustering while reducing the number of clusters to preserve $N_+$ and $N_-$, we find the inertial range can be extended. In Figure \ref{100clustering} (c) samples consists of $n_c=20$ vortices in each cluster, distributed between $r_{\rm min}=\xi^{-1}$ and $r_{\rm max}=100\pi\xi$, with $\bar\alpha=0.75$, and giving $\sim 2$ decades of inertial range. We note that, compared with Fig.~\ref{randAll}, $\bar k$, shown in Fig.~\ref{100clustering} (b) and (c) does not correspond so well with the lower bound on the inertial range, presumably because of the significant space between clusters in the neutral system.


\section{Forced 2D quantum turbulence}
\subsection{Scenario of forced turbulence\\ in a compressible 2D superfluid}
The canonical model of 2D classical turbulence consists of a velocity field described by the 2D Navier-Stokes equation
\begin{equation}\label{2dns}
\frac{\partial \mathbf{v}}{\partial t}+(\mathbf{v}\cdot\nabla)\mathbf{v}=-\frac{1}{\rho}\nabla p+\nu\nabla^2\mathbf{v}-\lambda\mathbf{v}+\mathbf{f}_\mathbf{v}.
\end{equation}
\begin{figure}[!tb]
\begin{center}
\includegraphics[width=0.9\columnwidth]{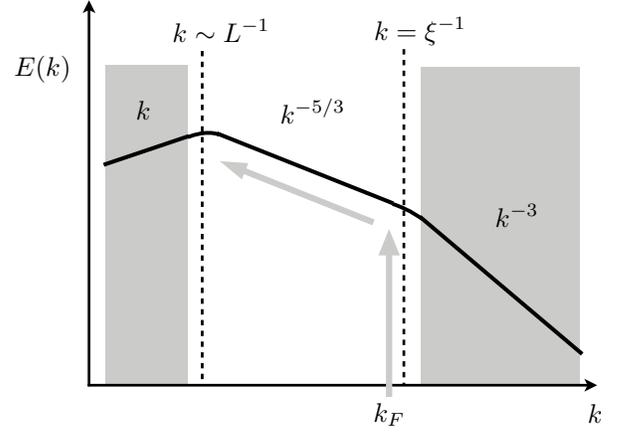}
\caption{ Illustration of an inertial range (non-shaded region) for the incompressible portion of kinetic energy in forced compressible 2DQT. The $E(k)\propto k^{-3}$ region arises from the structure of the vortex core and thus is not a signature of vortex configurations and vortex turbulence. This ultraviolet region can thus not support energy cascades, nor does this region correspond to enstrophy cascades. Net energy injected at $k_F\sim \xi^{-1}$ in the form of vortices can only move towards the infrared. The Kolmogorov law $E(k)\propto k^{-5/3}$ occurs in the inertial range of fully developed turbulence. The far-infrared region is given by $E(k)\propto k$ for a system with no net vorticity, and is evident for $k\ll L^{-1}$ where $L$ is the largest intervortex distance. For forcing at smaller wavenumbers, the spectrum may be more complex, possibly involving other forms of energy and enstrophy flux. 
\label{cascadeSchematic}}
\end{center}
\end{figure}
The density $\rho$ of the incompressible fluid is held constant by the pressure field $p$, $\nu$ is the kinematic viscosity, $\mathbf{f}_\mathbf{v}$ is a forcing term and $\lambda$ represents linear frictional damping arising from irreducible 3D aspects of the system in which the 2D flow resides. If the fluid is subjected to suitable forcing it will develop an inverse energy cascade and a direct enstrophy cascade, with associated $k^{-5/3}$ and $k^{-3}$ power laws respectively~\cite{Rutgers1998a}. An inverse energy cascade induced by small-scale forcing can be steady because the $-\lambda\mathbf{v}$ term damps energy at large length scales~\cite{Boffetta12a}. For a homogeneous compressible superfluid subject to forcing from an external potential, \eref{qnsdef} can be written as
\begin{equation}\label{qnsF}
\frac{\partial \mathbf{v}}{\partial t}+(\mathbf{v}\cdot\nabla)\mathbf{v}=-\frac{g}{m}\nabla \rho+\nu_q\nabla^2\mathbf{v}+\mathbf{f}_\mathbf{v},
\end{equation}
where the forcing $\mathbf{f}_\mathbf{v}\equiv -\nabla V(\pxb,t)/m$ is assumed to be spatially localized. The lack of a $-\lambda\mathbf{v}$ frictional damping term means, in the classical case, that if an inverse energy cascade develops as a result of steady forcing, it is not expected to be stationary. In the superfluid case, the compressibility of the fluid allows vortex-antivortex annihilation, which couples energy into the sound field. This interaction between the sound and vorticity fields renders the calculation of energy fluxes in compressible superfluid systems particularly difficult and somewhat ambiguous~\cite{Numasato10b}.

As we have shown above, in contrast with the classical Kraichnan scenario utilizing a 2D Navier-Stokes analysis, a $k^{-3}$ spectrum for $k > \xi^{-1}$ for a 2D quantum fluid is not caused by a direct enstrophy cascade but is rather a consequence of the vortex core structure, and thus should not be interpreted in terms of vortex configuration dynamics (note that vortex core shape excitations can be neglected, since they constitute a component of the sound field). Given forcing at a wavenumber $k_F\sim \xi^{-1}$, and minimal vortex-antivortex annihilation, the incompressible kinetic energy can only move toward the infrared. This scenario is shown schematically in Figure~\ref{cascadeSchematic}.

It has been shown that dipole recombination provides a route for a direct energy cascade to develop in 2D GPE dynamics~\cite{Numasato10b}. This mechanism provides a means for opposite-sign vortices to approach zero distance, coupling vortex energy to the sound field during vortex annihilation. However, if the forcing leads to significant clustering of like-sign vortices faster than recombination occurs, or prior to recombination occurring, dipole-decay will be strongly inhibited. This suggests that under the right conditions of forcing an inverse energy cascade can become the dominant mechanism of energy transport between distinct length scales.

\subsection{Kolmogorov constant and clustered fraction}
By making use of the universal onstrophy and the condition of continuity at $k\xi\approx 1$ we have found that the $k^{-5/3}$ power-law given by \eref{ir} describes the spectrum of numerically sampled vortex configurations that exhibit a $s^{-1/3}$ power law for the vortex separation data. While individual spectra and configurations do not give information about dynamics, in particular, the direction in $k$-space of any energy cascades, the power law  suggests the existence of an inertial range comprised of vortices. In a cascade, such a configuration will transfer incompressible energy between scales while conserving energy. Assuming the infrared portion of our double-power-law analysis, namely \eref{ir}, will also describe such a cascade, we can cast it as a statement about the Kolmogorov constant in terms of the one-vortex onstrophy and the slope of the radial wavefunction at the vortex core. 

To write \eref{ir} in standard form, we introduce the unique $N$-vortex quantity with dimensions energy/mass/time that can be constructed from $\Omega\xi^2$, $m$, and $\hbar/\Omega\xi^2$:
\EQ{\label{Erate}
\epsilon_N\equiv \frac{(\Omega\xi^2)^2}{m\hbar}N^{3/2}.
}
We then find
\EQ{\label{KolmogStan}
\frac{E_{i,{\rm C}}^N(k)}{m}=\bar{C}_{2D}\epsilon_N^{2/3}k^{-5/3}
}
where the remaining quantities have been absorbed into the dimensionless Kolmogorov constant:
\EQ{\label{Ck}
\bar{C}_{2D}\equiv \Lambda^{2} \left(\frac{\mu}{\Omega\xi^2}\right)^{1/3},
}
$\mu=\hbar^2/m\xi^2$ in the homogeneous system, and the bar notation distinguishes the quantum system. 
In the dilute Bose gas the 2D interaction parameter is $\mu/n_0=g_{2}=4\pi\hbar^2a/ml$, where $l$ is the characteristic thickness of the three-dimensional system~\cite{Numasato10b,Aioi2011}. In terms of this length we find
\EQ{\label{Cklength}
\bar{C}_{2D}=\Lambda^{2} \left(\frac{mg_{2}}{2\pi\hbar^2}\right)^{1/3}=\Lambda^{2} \left(\frac{2a}{l}\right)^{1/3}.
}
\begin{figure*}[!t]
\begin{center}
\includegraphics[width=\textwidth]{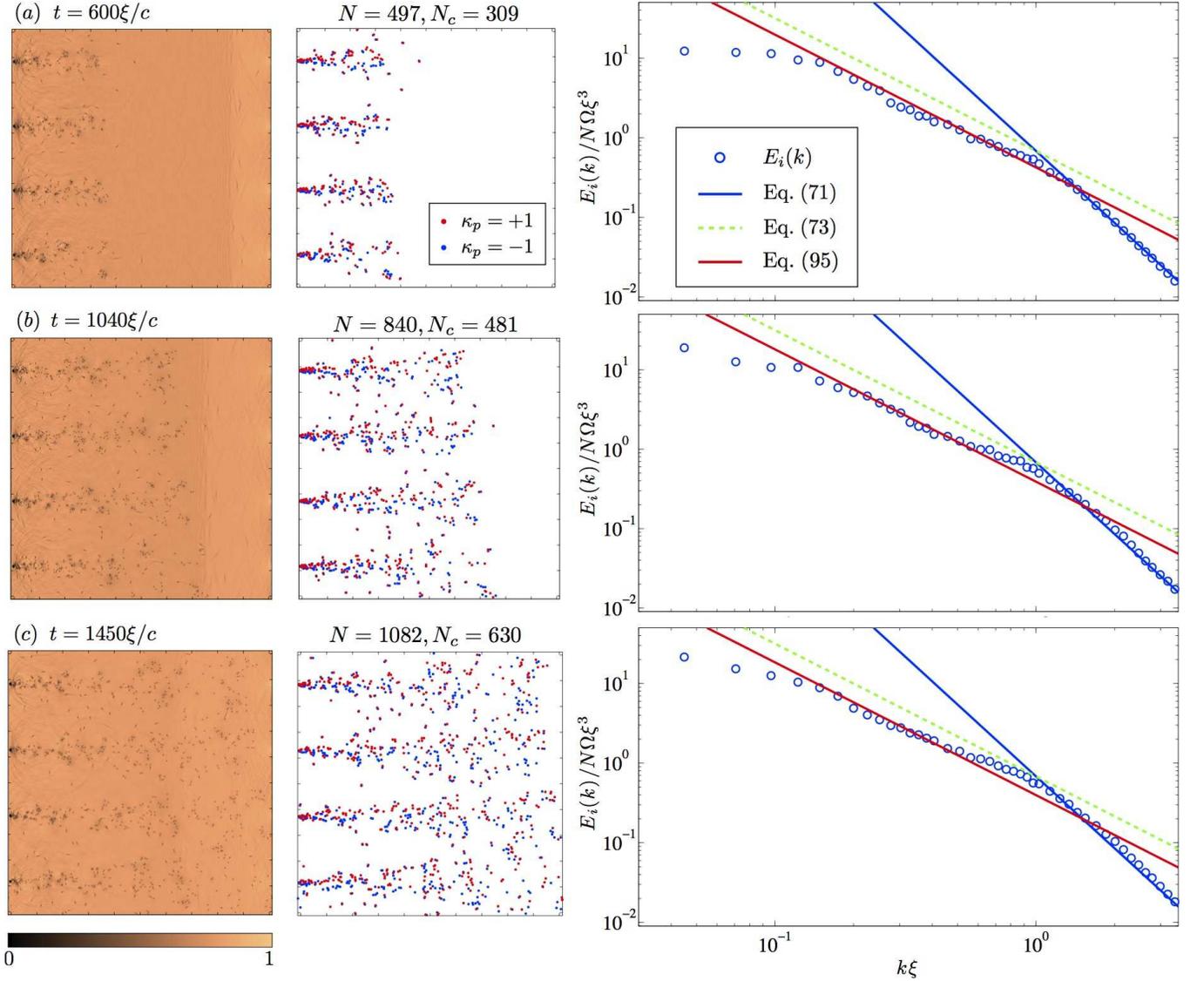}
\caption{Time evolution of grid turbulence in damped GPE.  Left: particle density (rescaled to the peak density). Center: vortices colored by charge, with total ($N$) and clustered ($N_c$) numbers of vortices. The field of view is $(1024\xi)^2$. Right: Incompressible energy spectra (circles), with the Kolmogorov ansatz [red line, Eq. \eref{irgen}], the ansatz for a polarized cluster of $N$ vortices [dashed line, Eq.~\eref{ir}], and the universal $k^{-3}$ region [blue line, Eq.~\eref{uv}].\label{timeSeq}}
\end{center}
\end{figure*}
\begin{figure}[!t]
\begin{center}
\includegraphics[width=1\columnwidth]{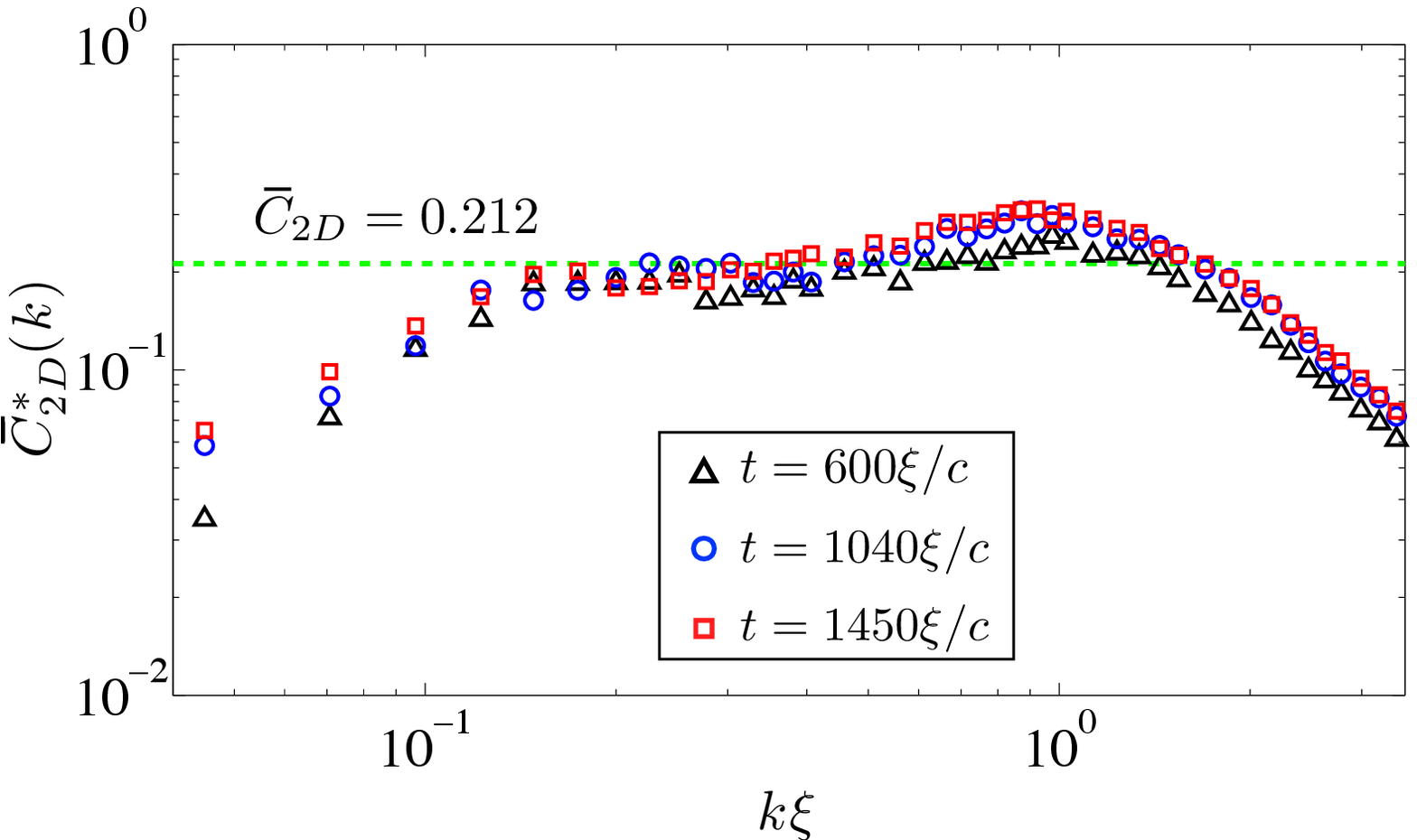}
\caption{Plot of $\bar{C}^*_{2D}(k)$ [see Eq.~\eref{k41}] computed from the grid turbulence kinetic energy spectra of Figure \ref{timeSeq}. The horizontal dashed line gives $\bar{C}_{2D} $ from Eq.~\eref{Cklength}. \label{Ckfig}}
\end{center}
\end{figure}
We emphasize that the physical input needed to arrive at this form of the Kolmogorov constant is (i) accounting for the structure of a compressible quantum vortex in determining the ultraviolet spectrum, and (ii) imposing continuity of the ultraviolet spectrum at $k\xi=1$ to a Kolmogorov power-law in the infrared. In classical turbulence, $C_{2D}\simeq 7$~\cite{Sreenivasan1995a}.
To give an example of how $\bar{C}_{2D}$ may be evaluated for a compressible superfluid exhibiting 2DQT, we consider a $^{87}$Rb BEC that is homogeneous in the $x-y$ plane, and that is harmonically trapped in the $z$-dimension with trap frequency $\omega_z=2\pi \times 5000$ Hz. We use $m=1.44\times 10^{-25} {\rm kg}$, $a=5.8{\rm nm}$, for which $l=\sqrt{2\pi\hbar/m\omega_z}=0.38$ $\mu$m, and $g_2=0.197 \hbar^2/m$.  For these values, $\bar{C}_{2D} = 0.212$.  Note also that by defining the configurational rate constant \eref{Erate}, we have confined this discussion to a scale invariant distribution involving $N$ vortices. This expression suggests that such a configuration can support an inverse energy cascade at the rate $\epsilon_N$.
\par
The foregoing discussion involves an ideal distribution of $N$ vortices configured with the $\alpha=1/3$ power law. It is clear from Fig.~\ref{100clustering} that the vortices do not have to all have the same sign of circulation, but they must be configured into clusters of vortices with the same sign. The universality of $C_{2D} $ in classical turbulence in incompressible fluids leads us to postulate that the condition that all $N$ vortices are power-law clustered can be further relaxed.  For fully developed quantum turbulence involving $N$ vortices, we interpret $N$ as the {\em participation number}, representing the number of vortices in a scale-free turbulent configuration, which in the case of a fully polarized cluster is maximal. Imperfect clustering involves fewer vortices in power-law cluster configurations, and an effective participation number that is the number of clustered vortices $N_c<N$, namely, the number with nearest-neighbors of the same sign. Making the replacement $N\to N_c$ in \eref{ir}, we propose the ansatz spectrum
\begin{equation}\label{irgen}
E_{i,{\rm C}}^N(k)=\Lambda^{2}\frac{N_c\Omega\xi^{3}}{(k\xi)^{5/3}},
\end{equation}
as a more general definition for systems that have incomplete clustering in the inertial scale range.
We test this hypothesis in the next section in dynamical simulations of the forced dGPE.  

It is important to note that the condition of continuity at $k\xi=1$ is no longer exactly met, since only the value $N_c\equiv N$ will produce an infrared spectrum with $k^{-5/3}$ that is continuous with the ultraviolet power law approximation at $k=1/\xi$. We also note that a more general measure of clustering, namely the {\em polarization index} ($P$) was introduced in Ref.~\cite{Lvov2007a} measuring the degree and type of spatial clustering of like-sign vortices in 3DCT. The Kolmogorov $k^{-5/3}$ spectrum was found to correspond to the partially polarized value $P=1/3$, while other scaling laws yield differing polarizations. The clustered fraction used in the present work is a simpler (global) measure of polarization as it does not contain information about the spatial distribution of vortices,  but it is only relevant for 2DQT.

We can also write down an expression for $\bar{C}_{2D}$ for a general energy spectrum that may be computed numerically from simulation data, by making use of the ansatz \eref{irgen}. This is equivalent to using \eref{KolmogStan} with $N\to N_c$, from which we can define the function
\begin{equation}\label{k41}
\bar{C}_{2D}^*(k)=E_{i,{\rm C}}^N(k)\frac{(k\xi)^{5/3}}{N_c\Omega\xi^3}\left(\frac{\mu}{\Omega\xi^2}\right)^{1/3}.
\end{equation}
In a region where the spectrum is approximately $k^{-5/3}$, $\bar{C}_{2D}^*(k)$ will be approximately constant, and it may be compared with the prediction \eref{Ck}. We test this numerically in the next section.

We note that in 3DQT an energy bottleneck has been predicted for the direct energy cascade~\cite{Lvov2007a,Lvov2008a}, and also observed in GPE simulations~\cite{Sasa2011a}. It occurs due a mismatch between the rates of energy transport at large length scales (hydrodynamic regime) and small length scales (Kelvin wave cascade). The mismatch causes energy to pile up at the length scale where the two cascades meet. This raises the possibility of a bottleneck in 2D, although Kelvin waves are disallowed in 2DQT, so this particular mechanism would not be relevant. However, for a given forcing mechanism, it is possible that the rate of transporting energy to large length scales may not be high enough to remove all of the vortex energy introduced at the forcing scale. Thus a bottleneck could still occur, and our assumption of continuity of the spectrum at $k\xi\simeq 1$ may not hold in general. We return to this question in the next subsection, where we find some indication of an energy bottleneck at the forcing scale in numerical simulations.

\subsection{Damped Gross-Pitaevskii Dynamics}\label{gpedyn}

We now consider a simulation of the forced dGPE that generates significant clustering of vortices of the same sign. 
The system consists of a homogeneous superfluid with periodic boundary conditions, stirred by dragging four Gaussian obstacle beams through it at a constant speed~\cite{Neely10a}, thus modeling grid turbulence in a BEC. When an obstacle is dragged through a superfluid sufficiently rapidly, superfluidity cannot be maintained. For slowly moving obstacles, the superfluid will adapt to the forcing, and vortices are not formed. Above a critical velocity $\mathrm{v}_c$~\cite{Frisch92a} vortex dipoles are periodically formed in the wake of the obstacle, injecting linear momentum into the superfluid. Sufficiently rapid motion ($\mathrm{v}\gg \mathrm{v}_c$) causes many vortices to be nucleated behind the obstacles in a chaotic fashion~\cite{Sasaki2010}, involving clustering of like sign vortices. Our choice of obstacle speed puts the system dynamics in the latter category. 

We work in units of $\mu$, $\xi$, and $\xi/c$ for energy, length, and time, respectively. In these units the specific parameters we choose (see the previous subsection) are $g_2=0.197\mu\xi^2$, corresponding to homogeneous density $n_0=\mu/g_2=5.26\xi^{-2}$, and $N_{tot}=5.5\times 10^6$ particles in a homogeneous 2D system of side length $L=1024\xi$. The Gaussian potentials each have fixed $1/e^2$ width of $w_0=\sqrt{8}\xi$, and height $V_0=100\mu$, and are initially located at $x=-L/2+8\xi$, $y=\pm L/8$, $\pm 3 L/8$. Numerically, we proceed by first finding a ground state of \eref{eom}, for a homogeneous system with periodic boundary conditions, subject to the localized obstacle beams. We then transform into a frame translating at ${\rm v}_0=0.8 c$, and maintain the obstacle locations relative to this frame, creating a dragging grid of stirring beams. A small amount of initial noise is added to the wave function to break the reflection symmetries of the system. We thus evolve the system according to \eref{eom} for the same potential, but with the Galilean transformed nonlinear operator ${\cal L}\to {\cal L}+i\hbar{\rm v}_0\partial_x$. During evolution the dimensionless damping rate is set to $\gamma=0.003$.  
\par
The time evolution of the system is shown in Figure \ref{timeSeq} at three times, at approximately $(1/3, 2/3,1)L/{\rm v}_0$, so as to avoid any periodic flow effects in the $x$-direction. The four obstacle beams generate many vortices (up to $N\sim 10^3$), and significant clustering ($N_c/N \gtrsim 0.6$). The incompressible kinetic energy spectrum shows a wide region that is well described by the $k^{-5/3}$ form of Equation \eref{irgen}. For later times [Figure \ref{timeSeq} (b), (c)] the spectrum shows a significant pile up around the forcing scale $k_F\sim \xi^{-1}$, suggesting a mismatch between the rates of injection and transport of incompressible kinetic energy. 
\par
In Figure \ref{Ckfig} we compare the function $\bar{C}^*_{2D}(k)$ [Eq.~\eref{k41}], as numerically computed from our simulation data, with the analytical prediction of the Kolmogorov constant $\bar{C}_{2D}$ [Eq.~\eref{Cklength}]. The region of $k^{-5/3}$ appears as a broad flat region that is in close agreement with $\bar{C}_{2D}=0.212$ pertaining to our simulation parameters.
\section{Conclusions}

To summarize, we have investigated relationships between the concepts of 2D turbulence in classical fluids and the emerging topic of 2D quantum turbulence of vortices, specifically as it relates to Bose-Einstein condensates. We established a link between the hydrodynamic limit of the damped GPE and the Navier-Stokes equations, providing an estimate of a quantum Reynolds number for superfluid flows in BECs. We have given a theoretical treatment of the incompressible kinetic energy spectrum that explicitly incorporates the vortex core structure in a compressible superfluid. The incompressible kinetic energy spectrum for a compressible superfluid is deconstructed in terms of single-vortex contributions determining a unique ultraviolet power-law where the energy spectrum scales as $k^{-3}$, and a contribution that depends on the configuration of vortices within the fluid that determines the infrared region of the spectrum. For the configurational regime we find:
\begin{enumerate}
\item The spectrum only depends on the distribution of vortex separations and the sign of the circulation of each quantum vortex. If the distribution of vortex separation $s$ for a system of vortices of the {\em same sign} is a power-law $\propto s^{-\alpha}$ with exponent $\alpha=1/3$, the kinetic energy spectrum will take the universal  Kolmogorov form $\propto k^{-5/3}$, as shown for point vortices~\cite{Novikov1975}. {\em Localized} clusters of $N$ vortices of the same circulation with this power law distribution can be constructed by sampling using a specific radial exponent $\bar{\alpha}$ that depends on the number of vortices and the scale range over which they are distributed.
\item The azimuthal velocity field of a large cluster is determined by $\bar\alpha$. By inflating the scale range of a cluster we find that $\bar\alpha$ increases, the velocity field is steepened and the inertial range expands to larger scales. In a neutral system the inertial range can be extended by increasing the size of clusters while decreasing their number.
\item The universal form of the UV region of the kinetic energy spectrum imposes a strong constraint. If the Kolmogorov power law occurs in the infrared region, then the postulate of continuity between the infrared and ultraviolet regions completely determines the spectrum when the ultraviolet and infrared regions are approximated as power laws. Physically, this corresponds to the inertial range extending down to the smallest configurational scale of the system $\sim \xi$.  We note that the postulate of continuity may not be relevant for all systems or forcing mechanisms.
\item We infer an analytical value for the Kolmogorov constant [Eq.~\eref{Ck}] under the conditions of spectral continuity at the cross-over scale for a system of vortices of the same sign. To assess the validity of this inference for dynamical situations we compare our analytical results with spectra from a numerical simulation of the forced dGPE for the specific case of a dragging a grid of obstacles through an otherwise homogeneous BEC. We find reasonable agreement provided we introduce the concept of a {\em clustered fraction} $N_c/N\leq1$, which is the fraction of vortices that have same-sign nearest neighbors. This measure discounts all vortex dipoles from the configurational analysis. We then observe good agreement between our Kolmogorov ansatz [Eq.~\eref{irgen}], and the spectrum calculated from the dGPE data. We also find that the predicted value of the Kolmogorov constant is in close agreement with the numerical simulations [Figure \ref{Ckfig}].
\end{enumerate}
We note that while our analysis indicates that vortex positions and circulations are enough to determine an approximate incompressible kinetic energy spectrum, the reverse is not necessarily true: a Kolmogorov spectrum does not carry information about specific vortex distributions.  Nevertheless, our analysis does indicate that the number of vortices in a quantum fluid can in principle be directly determined from the ultraviolet energy spectrum.  Moreover, the concept of a cascade in turbulence implies system dynamics and energy transport, yet aside from our numerical simulation example, our analytical approach is an instantaneous measure.  Importantly, one must determine means of characterizing vortex motion and relate such dynamics to the cascade concept.

The field of 2D quantum vortex turbulence is relatively new, compared with the much longer histories of 3D superfluid turbulence, 2D classical turbulence, and even dilute-gas Bose-Einstein condensation.  Point-vortex models have been extensively used in descriptions of superfluid dynamics as well as in 2D classical turbulence, although point-vortex distributions can only serve as approximate models of real 2D classical flows.  Our approach merges concepts from each of the above subjects in order to develop a new understanding of 2D quantum turbulence.  By considering the compressibility of a dilute-gas BEC, we find an analytical expression for the ultraviolet incompressible kinetic energy spectrum and an $N$-vortex equivalent of enstrophy in a quantum fluid, which we term the onstrophy. For the infrared region, point vortex models are sufficient, and vortex configurations serve to identify spectra as summarized above.  Taken together, the primary new outcome of our analysis is a link between vortex distributions, vortex core structure, and power-law spectra for 2D compressible quantum fluids.

Future work on 2D quantum vortex turbulence will involve numerical simulations and comparisons with our analytical results, extension of this analysis to confined and inhomogeneous density distributions, characterization of vortex dynamics and the time-dependence of energy spectra particularly in relation to clustering~\cite{White2012a}, inverse-energy cascades, and the nonthermal fixed point~\cite{Nowak2012a,Schole2012a}, and investigation of connections with weak-wave turbulence in BEC~\cite{Zakharov1992,Dyachenko1992,Berloff02a,Nazarenko06a,Proment09a,Kozik09a,Nowak2011a}.  We also believe that observing vortex distributions such as the $\alpha=1/3$ power-law for localized clusters may provide a new means of quantitatively characterizing 2D quantum vortex turbulence through direct experimental observations of vortex locations in a forced 2D superfluid, and we are working towards realizing such experimental observations.

\acknowledgements
We thank Matt Reeves for assistance with numerical simulations, and Aaron Clauset for providing code for sampling power-law distributions.
We also thank Thomas Gasenzer, Gary Williams, Matt Reeves, Sam Rooney, Blair Blakie, Murray Holland, Giorgio Krstulovic, Michikazu Kobayashi, and Ewan Wright for useful discussions.
\par
We are supported by the Marsden Fund of New Zealand (contract UOO162) and The Royal Society of New Zealand (contract UOO004) (AB), and the US National Science Foundation grant PHY-0855467 (BA).


\begin{thebibliography}{10}
\bibitem{whatisenstrophy}
In two-dimensions, enstrophy $\int dx\, dy\, \omega^2$ is a measure of the
  vorticity $\omega = [ \nabla \times v ]_z$ about the $z$ axis, where $v$ is
  the fluid velocity in the plane normal to $z$.
\end{thebibliography}

\end{document}